\documentclass[journal=jacsat,manuscript=article]{achemso}

\usepackage{physics}
\usepackage{amssymb}
\usepackage{amsmath}
\usepackage{bm}
\usepackage{lineno}

\usepackage{xr-hyper}
\makeatletter
\newcommand*{\addFileDependency}[1]{
\typeout{(#1)}
\@addtofilelist{#1}
\IfFileExists{#1}{}{\typeout{No file #1.}}
}\makeatother

\newcommand*{\myexternaldocument}[1]{%
\externaldocument{#1}%
\addFileDependency{#1.tex}%
\addFileDependency{#1.aux}%
}

\myexternaldocument{2_ESI}

\usepackage[hidelinks]{hyperref}
\hypersetup{
    colorlinks=False,
    linkcolor=blue,
    filecolor=blue,      
    urlcolor=black,
    pdftitle={Overleaf Example},
    }
\usepackage[version=3]{mhchem} 

\usepackage{graphicx}
\usepackage{mathtools}
\usepackage{siunitx}
\usepackage{xcolor}

\newcommand{\RomanNumeralCaps}[1]
\linenumbers

\DeclarePairedDelimiterXPP\BigOSI[2]%
  {\mathcal{O}}{(}{)}{}%
  {\SI{#1}{#2}}

\title{Performance enhancement of electrocatalytic hydrogen evolution through coalescence-induced bubble dynamics}

\author{Aleksandr Bashkatov}
\affiliation[Twente]{Physics of Fluids Group, Max Planck Center for Complex Fluid Dynamics and J. M. Burgers Centre for Fluid Dynamics, University of Twente, Enschede, Netherlands}
\email{a.bashkatov@utwente.nl}
\author{Sunghak Park}
\affiliation[Leiden]{Leiden Institute of Chemistry, Leiden University, Leiden, Netherlands}
\author{Çayan Demirkır}
\affiliation[Twente]{Physics of Fluids Group, Max Planck Center for Complex Fluid Dynamics and J. M. Burgers Centre for Fluid Dynamics, University of Twente, Enschede, Netherlands}
\author{Jeffery A. Wood}
\affiliation[Twente2]{Soft Matter, Fluidics and Interfaces, MESA+ Institute for Nanotechnology, J. M. Burgers Centre for Fluid Dynamics, University of Twente, Enschede, Netherlands}
\author{Marc T.M. Koper}
\affiliation[Leiden]{Leiden Institute of Chemistry, Leiden University, Leiden, Netherlands}
\author{Detlef Lohse}
\affiliation[Twente]{Physics of Fluids Group, Max Planck Center for Complex Fluid Dynamics and J. M. Burgers Centre for Fluid Dynamics, University of Twente, Enschede, Netherlands}
\alsoaffiliation[MaxPlanck]{Max Planck Institute for Dynamics and Self-Organization, G\"{o}ttingen, Germany}
\author{Dominik Krug}
\affiliation[Twente]{Physics of Fluids Group, Max Planck Center for Complex Fluid Dynamics and J. M. Burgers Centre for Fluid Dynamics, University of Twente, Enschede, Netherlands}
\email{d.j.krug@utwente.nl}

\begin{document}
\begin{abstract}
The evolution of electrogenerated gas bubbles during water electrolysis can significantly hamper the overall process efficiency. Promoting the departure of electrochemically generated bubbles during (water) electrolysis is therefore beneficial. For a single bubble, a departure from the electrode surface occurs when buoyancy wins over the downward-acting forces (e.g. contact, Marangoni, and electric forces). In this work, the dynamics of a pair of H$_2$ bubbles produced during hydrogen evolution reaction in 0.5 M H$_2$SO$_4$ using dual platinum micro-electrode system is systematically studied by varying the electrode distance and the cathodic potential. By combining high-speed imaging and electrochemical analysis, we demonstrate the importance of bubble-bubble interactions for the departure process. We show that bubble coalescence may lead to substantially earlier bubble departure as compared to buoyancy effects alone, resulting in considerably higher reaction rates at constant potential.  However, due to continued mass input and conservation of momentum repeated coalescence events with bubbles close to the electrode may drive departed bubbles back to the surface beyond a critical current, which increases with the electrode spacing. The latter leads to the resumption of bubble growth near the electrode surface, followed by buoyancy-driven departure. While less favourable at small electrode spacing, this configuration proves to be very beneficial at larger separations increasing the mean current up to 2.4 times compared to a single electrode under the conditions explored in this study.
\end{abstract}

\section{Introduction}

Water electrolysis is likely to become a central technology in the CO$_2$-neutral energy system of the future. Apart from being a potential energy carrier and fuel, hydrogen gas serves as 
a feedstock for the chemical (e.g. ammonia production for fertilisers) and steel industries (coal replacement), and refineries (methanol, synthetic fuels) \cite{brandon2017clean,staffell2019role,dawood2020hydrogen}.
Yet, the process efficiency requires further improvement to compete on the energy market and enable large-scale hydrogen production. In both conventional alkaline and proton exchange membrane water electrolyzers a considerable part of the overpotentials and hence losses can be attributed to the formation of H$_2$ and O$_2$ bubbles, present at the electrodes and in the bulk.\cite{lee2020accelerating,swiegers2021prospects, yu2022tuning, yuan2023bubble} These bubbles block the electrodes by masking their active surface area, reducing the number of nucleation sites. Additionally, they raise ohmic resistance by blocking the ion-conducting pathways.\cite{zhao2019gas,angulo2020influence,angulo2022understanding} It is therefore vital to maintain a bubble-free electrode area for continuous catalytic activity. Enhanced removal of gas bubbles and deeper insights into their evolution processes will benefit further optimization of the system's energy efficiency \cite{shih2022water}.

Various methods have been developed to aid bubble departure, categorized as active (e.g., sonication, centrifugal forces, mechanical convection, pressure modulation, external magnetic fields) and passive approaches.\cite{he2023strategies, swiegers2021prospects, yuan2023bubble,gross2021mitigating} Passive methods, preferred for their energy-efficiency, primarily involve surface modifications to alter the wettability\cite{tang2022bioinspired} of the catalytic surface.\cite{krause2023hydrogen} For example, superhydrophilic surfaces facilitate earlier bubble departure due to the reduced contact angle at liquid-solid interfaces.\cite{nam2011single,lu2014ultrahigh,li2015under,hao2016superhydrophilic,Iwata_2021,andaveh2022superaerophobic,cheng2023bubble}

The bubble removal process can also benefit from hydrophobic surfaces. One example is the bubble-free electrolysis concept that employs a hydrophobic porous layer adjacent to a porous electrode. This prevents bubble formation within the catalyst, guiding the generated gas by capillary effects through the hydrophobic layer \cite{Winther_Jensen_2012,tiwari2019new,tsekouras2021insights,hodges2022high}.
A different concept to enhance gas removal, which was shown to hold promise based on theoretical analysis,\cite{Kadyk_2016} is the use of hydrophobic islands on the electrode as preferential nucleation sites. Also practically, this has been shown to be feasible using electrodes partially covered with hydrophobic spots made of polytetrafluoroethylene (PTFE)\cite{Teschke_1983,Teschke_1984,Brussieux_2011}. This allows to guide the produced gas away from the active areas of the electrode with the potential to lower the bubble-induced overpotentials \cite{Teschke_1983,Teschke_1984}. 
\citet{Brussieux_2011} demonstrated that, depending on the size of and distance between islets, parameters of the gas release such as bubble size and location can be controlled, but did not study the effect on electrode performance. More recently, \citet{Lake_2022} found that densely packed Pt-coated micropost arrays promote consistent release of smaller bubbles through their mutual coalescence. While this enhanced the stability of the current compared to untextured electrodes, it did not lead to performance gains when normalising by the active surface area in this system, due to additional bubbles forming in between the pillars.
In this context, coalescence induced removal of bubbles is of particular interest. Coalescence leads to a reduction in surface energy and this difference is in part converted to kinetic energy, causing the bubble to jump off the surface without having to rely on buoyancy.  This makes this removal process also highly attractive in microgravity applications.\cite{matsushima2003water,zhou2018modified,brinkert2018efficient, akay2022electrolysis,akay2022releasing, raza2023coalescence, bashkatov2021dynamics}

However, a detailed understanding of the mechanism and quantification of the extent to which coalescence-induced dynamics can be exploited to improve the performance of gas-evolving electrodes is still lacking. This also applies to parameter optimisation, which in view of complications such as a possible bubble return to the electrode surface,\cite{janssen1970effect, sides1985close, hashemi2019versatile,ikeda2023tert, westerheide1961isothermal,bashkatov2021dynamics,bashkatov2022} is highly nontrivial.
We address these open questions in the present work by studying the coalescence-driven dynamics of hydrogen bubbles produced at a dual micro-electrode during water electrolysis. This new setup allows precise control of important parameters such as the bubble size during coalescence, while also providing excellent observability of the dynamics. We demonstrate that coalescence events may lead to both premature bubble departure compared to buoyancy effects alone and the return of departed bubbles to the surface of the electrode, substantially altering the reaction rates. The dual micro-electrode configuration shows, depending on the applied potential and inter-electrode distance, up to a 2.4-fold increase in current compared to a single micro-electrode.

\section{Methods}
\begin{figure}\centering
	\includegraphics[width=0.5\textwidth]{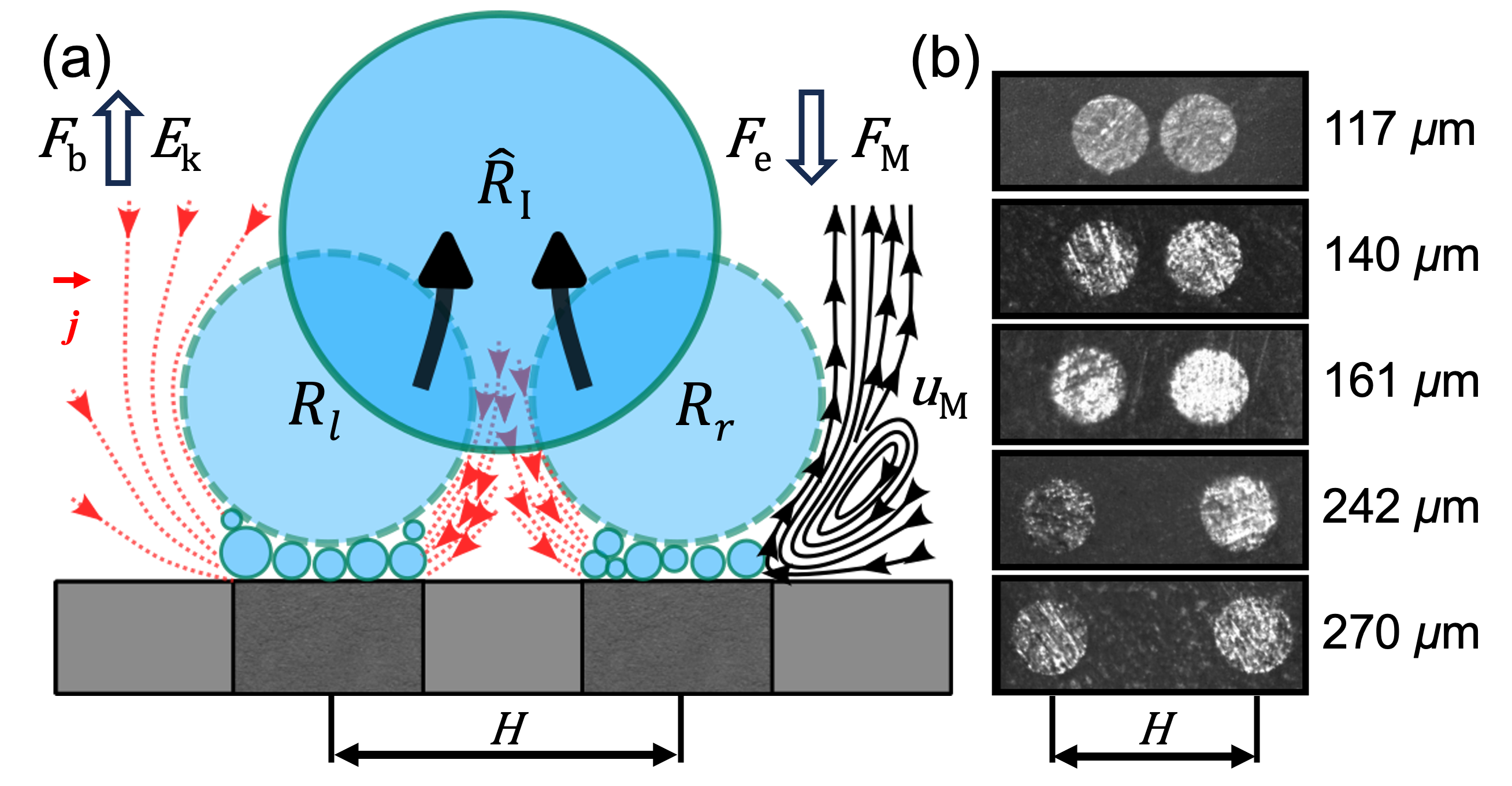}
	\caption{(a) Schematic of the dual micro-electrode and two H$_2$ bubbles sitting on the carpet of micro-bubbles. Each growing bubble is subject to a force balance including buoyancy, electric, and Marangoni forces. The red lines represent current density ($j$) and the black streamlines on the right represent the Marangoni convection with velocity $u_M$. $E_{k}$ is the kinetic energy released during the coalescence of the left ($R_l$) and right ($R_r$) bubbles. (b) Top view of the five dual micro-electrodes with different interelectrode distance (\textit{H}).
		\label{fig:scheme}}
\end{figure}

The pairs of H$_2$ bubbles sketched in figure \ref{fig:scheme}a were generated at the surface of a dual platinum micro-electrode during the hydrogen evolution reaction (HER). 
The experiment was performed in a three-electrode electrochemical cell filled with 0.5 M H$_2$SO$_4$.

The fabrication of dual micro-electrodes followed a previously established method.\cite{park2023solutal} Briefly, two Pt wires ($\varnothing$100 $\mu$m, 99.99\%, Goodfellow) were sealed into a soda-lime glass capillary (outer diameter $\varnothing$1.4 mm, inner diameter $\varnothing$1.12 mm, Hilgenberg) by gently softening the capillary in a flame. Five different values for the interelectrode distance \textit{H} were established and tested, as shown in fig.~\ref{fig:scheme}b. The electrode surface underwent electrochemical cleaning (potential cycling between 0.03...1.35 V vs. RHE , repeated 50 times) after being mechanically polished with sandpaper (2000 grit), sonicated and rinsed with ultrapure water.
The cell used here closely resembles that used in earlier studies \cite{yang2015dynamics,bashkatov2022,park2023solutal}. The dual micro-electrode (cathode) is inserted horizontally facing upward in the base of a cuboid glass cuvette (Hellma) with dimensions of 10 $\times$ 10 $\times$ 40 mm$^3$.   
 The system is completed by the reference electrode (Ag/AgCl) and counter electrode ($\varnothing$ 0.5 mm Pt wire) both inserted vertically from the top. The electrochemical cell is controlled by a potentiostat (BioLogic, VSP-300, 6 channels) at a constant potential of -0.2 to -2.8 V (vs. RHE). Each of the two electrodes is connected to and controlled by a separate channel of the potentiostat. For each experimental run, the electric current was recorded with a sampling rate of at least 1 kHz over a period of 30 s.
The optically transparent cell allows visualization of the bubble dynamics 
using a shadowgraphy system. It consists of LED illumination (SCHOTT, KL 2500) with a microscope, connected to a high-speed camera (Photron, FASTCAM NOVA S16), providing a spatial resolution of 996 pix/mm. Image recording was typically performed at 5 kHz, unless otherwise stated.  High-speed recording up to 264 kHz was used to resolve individual coalescence events. The bubble radius was extracted by standard image processing routine based on the Canny edge detection method in Matlab R2022b (for further details see Supplemental Material in \citet{bashkatov2019}). To measure the velocity fields around H$_2$ bubbles presented in figure \ref{fig:dual_sum}, monodisperse polystyrene particles (microParticles GmbH) of $\varnothing$5 $\mu$m were seeded into the electrolyte. These particles are neutrally buoyant with a mass density of 1.05 g/cm³. The resulting series of images, recorded at 1000 frames per second, were processed by the software DaViS 10, which employs a Particle Tracking Velocimetry (PTV) algorithm to track each particle over 25 ms shortly before departure. Due to the limited number of particles close to or at the bubble-electrolyte interface, the resulting tracks of the particles were collected for 60 bubbles. Subsequently, the tracks were converted into a vector field using a binning function that interpolates local tracks on a specified fine grid.

\section{Single electrode}
\label{sec:single}
\begin{figure*}[h!]\centering
	\includegraphics[width=0.7\textwidth]{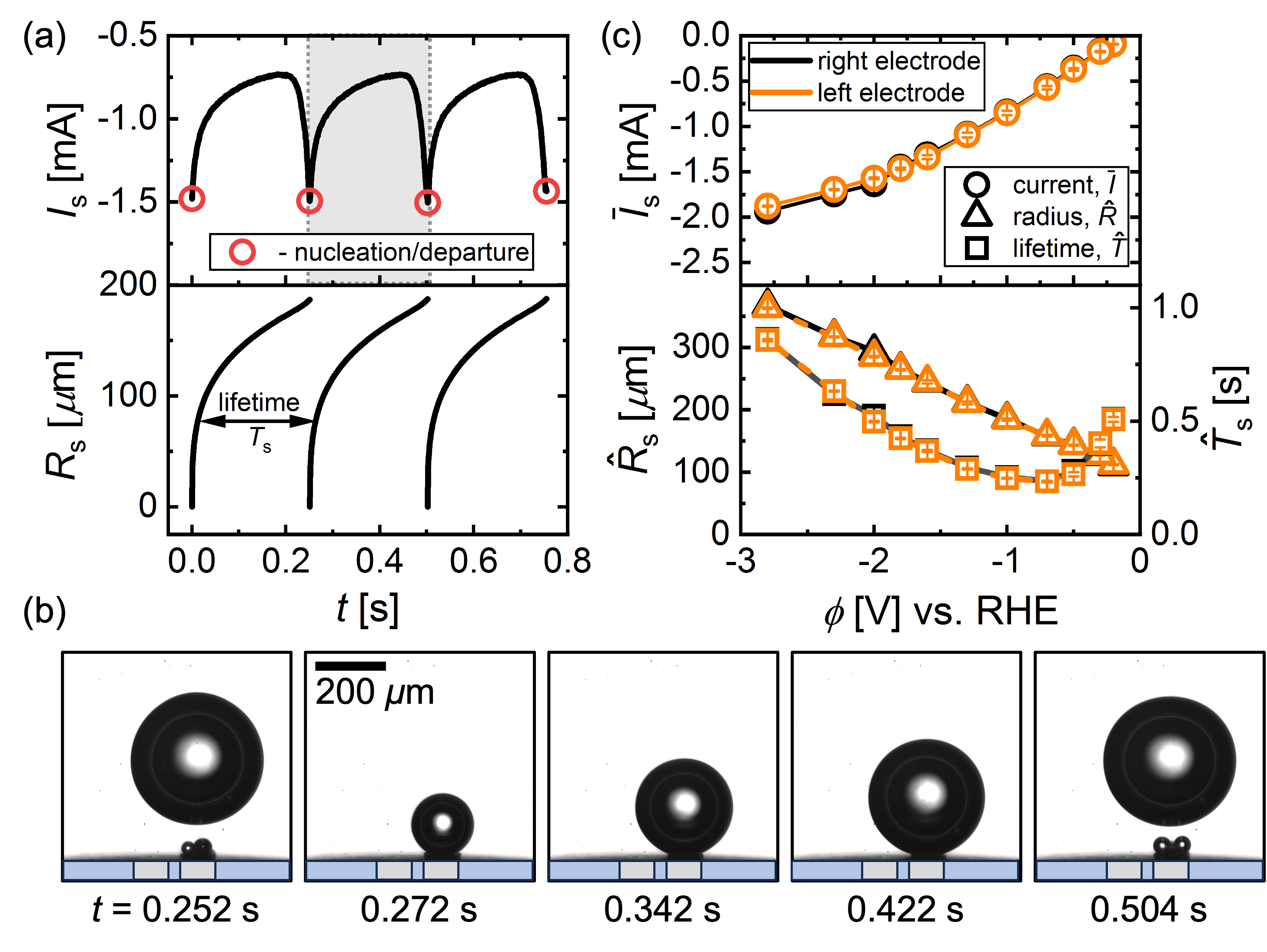}
	\caption{(a) The electric current and radius over time representing three complete cycles of bubble evolution at $\phi = -1.0$ V at a single micro-electrode. The red circles mark nucleation and departure instants of time. (b) Shadowgraphs displaying the evolution cycle, marked in grey in (a). (c) The averaged electric current (circles), departure radius (triangles) and lifetime (squares) versus the potential for the right (black) and left (orange) electrodes, run separately. 
 Image recording performed at 500 frames/second.
		\label{fig:sin}}
\end{figure*}

To set the baseline, we briefly report the results for the case where only a single electrode is operated, which has been studied previously.\cite{kristof1997effect,brandon1985growth,fernandez2014bubble,yang2015dynamics,massing2019,hossain2020thermocapillary,bashkatov2019,bashkatov2021dynamics,bashkatov2022,babich2023situ,zhan2023dynamics,meulenbroek2021competing} 
As an example, figure \ref{fig:sin}a shows the transient current ($I_{s}$) and radius ($R_{s}$) of the bubble for three complete bubble evolution cycles at -1.0~V. Shadowgraphs corresponding to a complete cycle from nucleation \cite{perez2019mechanisms,chen2015electrochemical,german2018critical} to departure are included in figure \ref{fig:sin}b. This process is highly periodic with a bubble lifetime $T_{s}$. The evolution of the bubble has a strong influence on the reaction current, for which the maxima in cathodic current marked by the red circles coincide with the departure of the bubble. This is immediately followed by the nucleation of a new bubble, whose growth in the vicinity of the electrode leads to a considerable reduction in $I_{s}$ of up to 50\% in this case. This continues until the next bubble departure, after which the cycle repeats itself. 

Finally, figure \ref{fig:sin}c summarizes how the average electric current $\overline{I}_{s}$, where the overline denotes an average over $t$, the departure radius ($\hat{R}_{s}$), and lifetime ($\hat{T}_s$) varies for different cathodic potentials ($\phi$). All statistics are averaged over multiple bubble cycles with error bars representing the standard deviation. The figure also confirms that consistent results are obtained from both electrodes. 
 
In this system, bubble formation occurs already at low overpotentials. Micron-sized bubbles form on the electrode surface and continuously coalesce to form a single larger bubble. This larger bubble is typically not in direct contact with the electrode surface, but rather resides on the layer of microbubbles\cite{bashkatov2022}. It continues to grow via the intensive coalescence with these microbubbles and via gas diffusion\cite{sepahi2023mass}. In this case, departure of the bubble occurs once the retaining forces due to the electric field\cite{bashkatov2022}, thermal\cite{young1959motion,guelcher1998thermocapillary,lubetkin2003thermal,yang2018marangoni} and solutal\cite{park2023solutal} Marangoni effects are overcome by buoyancy (see figure \ref{fig:scheme}a). The thermal Marangoni effect is related to the Joule heating caused by the locally high current density ($j$) at the bubble foot as indicated in figure \ref{fig:scheme}a. The effect therefore scales (via Ohm's law) with $j^2$ and prevails at high overpotentials. The solutal effect due to the depletion of the electrolyte at the electrode is expected to depend linearly on $j$ and therefore dominates at lower overpotentials ($\phi \gtrapprox -0.7$ V in the present case)\cite{park2023solutal}. The electric force is directly proportional to $\phi$ and therefore all retaining forces diminish as the overpotential is reduced, which explains the decreasing trend of the departure radius $\hat{R}_{s}$ as $|\phi|$ is made less negative. Since the bubble captures almost all the produced gas\cite{yang2015dynamics,park2023solutal}, the departure period follows from the time it takes to produce the gas contained in the bubble volume and $T_{s}$ is therefore proportional to $R_{s}^3/I$. 

\section{Dual electrode}
\subsection{Modes of bubbles evolution}
\begin{figure*}[h!]\centering
	\includegraphics[width=1\textwidth]{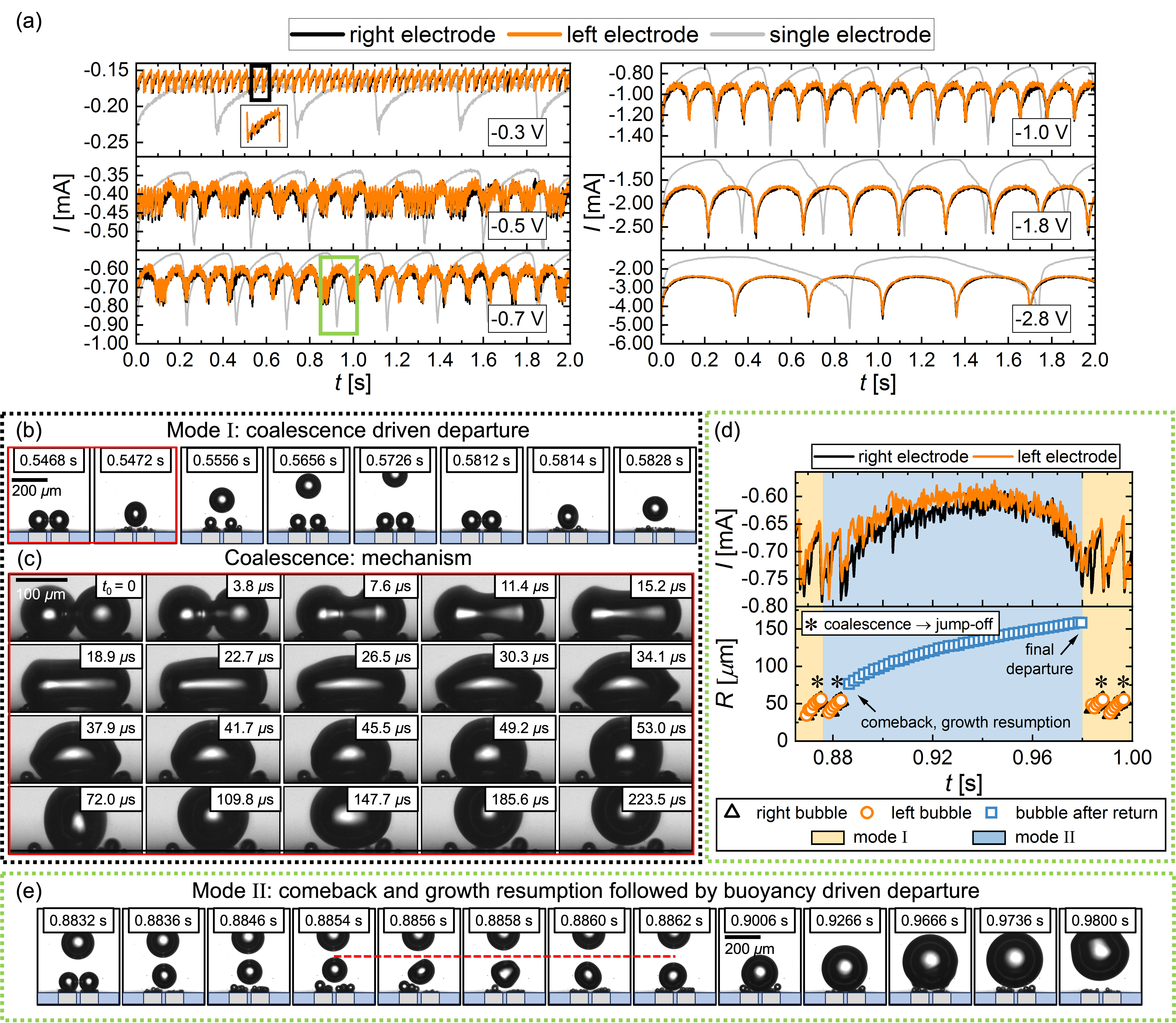}
	\caption{
 (a) Electric current over 2 seconds (out of 30 seconds) of the experimental run at various potentials ($\phi$). The black and orange curves represent the electric current measured at the right and left electrodes, respectively. Grey lines represent corresponding results for single electrode operation. (b) Snapshots depict the bubble evolution following mode I as indicated in (a) by the black rectangular inset at -0.3 V. (c) Snapshots detailing the coalescence-driven departure process recorded at -0.5 V. $t_0$ is one frame before the coalescence process begins.
(d) A zoomed-in view of the current at -0.7 V, shown by the green rectangle in (a), with corresponding evolution of $R(t)$. The orange and blue shades correspond to modes I and II, respectively. (e) Mode II of bubble evolution from (d). The red line indicates the maximum height reached by the departed bubble. Recordings in (b) and (e) were performed at 5 kHz, and at 264 kHz in (c).
		\label{fig:sc}}
\end{figure*}

From now on, both electrodes are operated  simultaneously, independently of each other, and at the same potential. Initially, we will only consider the pair with a separation of $H = 117$ $\mu$m. 
The measured currents for this configuration are plotted in figure \ref{fig:sc}a for different potentials. Time traces of the current for both electrodes ('left' and 'right') are included and for reference we also show the current signal measured when only a single electrode is operated at the same potential (grey line). Focusing initially on the lowest overpotential, $\phi = -0.3$ V, the current oscillations remain periodic during dual operation; however, both the period and amplitude are notably diminished. The reason for this can be understood from the corresponding shadowgraphs presented in figure \ref{fig:sc}b, which illustrate the bubble dynamics over one period (shown by black box in fig. \ref{fig:sc}a). 

Similar to what is observed for a single electrode, a larger bubble forms and grows initially at each of the two electrodes, leading to a gradual reduction in the current. This process continues until the two bubbles touch and coalesce, which is followed by the departure of the merged bubble along with a spike in the current (see inset at -0.3 V in fig. \ref{fig:sc}a). Figure \ref{fig:sc}c details this coalescence process, which happens on the order of micro-seconds, and the emerging deformations of the bubble shape. The coalescence induced jump-off is powered by the released surface energy\cite{soto2018coalescence, lv2021self,bashkatov2021dynamics}. While the majority of this energy is dissipated through the capillary waves seen in figure \ref{fig:sc}c \cite{sanjay2021bursting,raza2023coalescence}, the fraction that is transformed into kinetic energy (less than 1\%, for details see Supporting Information) can cause bubble departure at much smaller radii than in the buoyancy-driven scenario, for the newly formed bubble. Together with the fact that each of the coalescing bubbles only contributes half the volume, this explains the significantly enhanced departure frequency.    

At higher overpotential at $\phi = -0.5$ V, events with a much longer period length start appearing intermittently in the current traces. These events become more frequent and dominate the signal at $\phi = - 0.7$ V, before almost fully superseding  the high-frequency coalescence pattern at $\phi \leq -1.0$ V. 
In order to elucidate the underlying bubble dynamics, we provide an enlarged view of a segment of the current signal at $\phi= -0.7$ V (green box) in figure \ref{fig:sc}d along with the size evolution of the bubbles. The first bubble departure included in figure \ref{fig:sc}d proceeds analogously to the one shown in figure \ref{fig:sc}b, and the bubble continues to rise away from the electrode after the coalescence induced take-off. We will refer to this as 'mode I' from now on. However, as the corresponding shadowgraphs in figure \ref{fig:sc}e show, even though the bubble also jumps off after the second coalescence event, it is eventually brought back to the surface through repeated coalescence with newly formed bubbles at both electrodes (see period between $t = 0.8854$ s and $t = 0.8862$ s). Following this return, the bubble rests between the two electrodes just above the surface. There, it continues to grow until a buoyancy driven departure (at $R_{II} = 158\:\mu$m vs. $R_I = 72\:\mu$m), which explains an order of magnitude longer lifetime ($T_{II}$ = 104.4 ms vs. $T_I$ = 8.4 ms) of the bubble in this instance. We will denote this as 'comeback mode' or 'mode II'. 

\begin{figure}[h]\centering
	\includegraphics[width=0.45\textwidth]{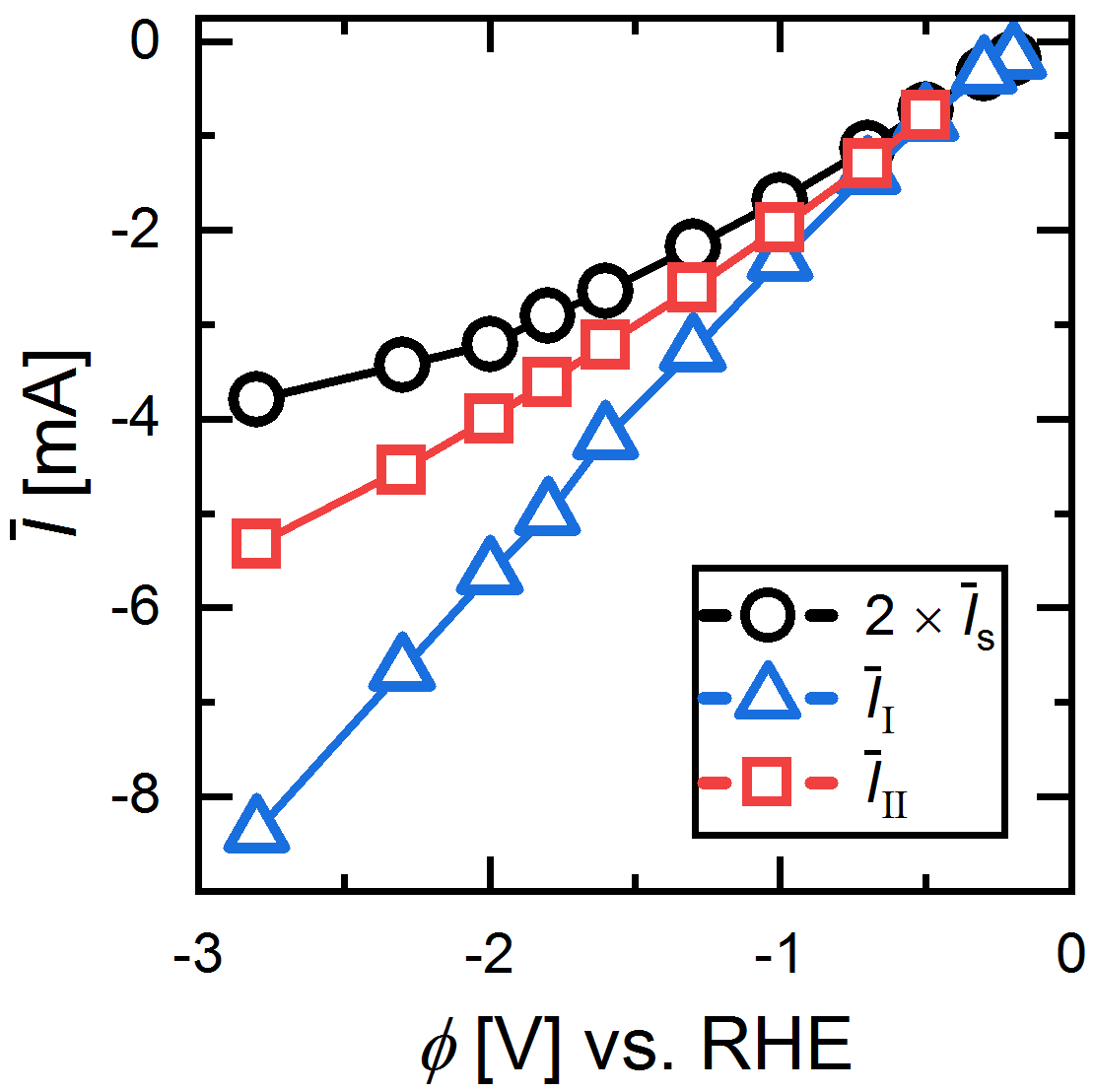}
	\caption{Electric current ($\overline{I}$) vs. potential ($\phi$) for single electrode (black) and for modes I (blue) and II (red) at dual micro-electrode. Both $\overline{I}_I$ and $\overline{I}_{II}$ are the sum of the currents at the left and right electrodes.
		\label{fig:sc_sum}}
\end{figure}

It is evident from figure \ref{fig:sc}a that the dynamics induced by coalescence have a strong impact not only on the current fluctuations, but also on the mean current at a specific potential. To analyse this, we compare period averaged currents for the two modes ($\overline{I}_I$ and $\overline{I}_{II}$, taken to be the sum of the currents at both electrodes) to $2\times \overline{I}_{s}$ in figure \ref{fig:sc_sum}. 
Note that it is possible to determine $I_I$ even at high potentials where mode II prevails by considering only the time until the first coalescence, leading to temporary departure of the bubble (see figure \ref{fig:sc}d). 
Despite the much faster gas removal, the current at low overpotentials ($\phi \gtrapprox - 0.7$ V) remains the same or even slightly decreases in dual operation compared to the single electrode case. This can be attributed to the additional shielding by the second bubble in the vicinity of the electrode and the diffusive competition between the two reaction sites, both of which lower performance. However, the benefits of the accelerated gas removal increasingly outweigh these effects as the overpotential is increased. This is particularly true for mode I, where the current is more than double than that of the single electrode at the same potential for the most negative values of $\phi$ investigated. While this clearly demonstrates the potential for performance enhancement through coalescence-induced gas removal, the effective performance enhancement is reduced to less than 50\% for the current electrode spacing due to the prevalence of bubble return (mode II) at higher overpotentials. The currents in mode II are consistently lower compared to mode I because the electrode separation is so small, that the returning bubble still blocks a large part of both electrodes (see figure \ref{fig:sc}e), even though it is located half-way between them. 

\subsection{Phase diagram}
\label{sec:phase}
\begin{figure*}[h!]\centering
	\includegraphics[width=1\textwidth]{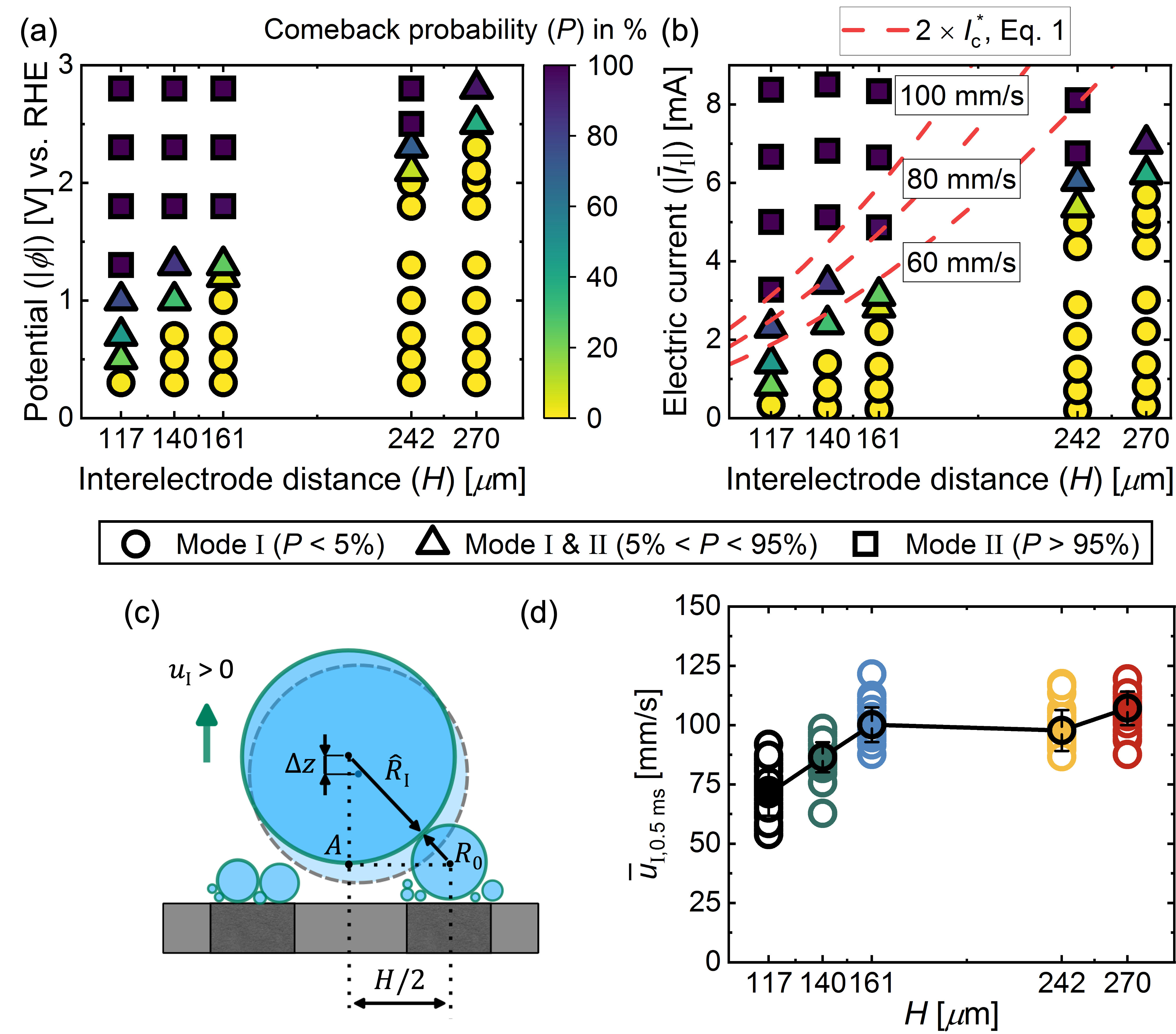}
	\caption{Phase diagram representing the probability (\textit{P}) of the bubble coming back after initial departure shown in terms of (a) Potential and (b) Current vs. $H$. The color bar scales the probability from 0 to 100\%. The circles denote \textit{Scenario I}, i.e. when $P$ is less than 5\%, and squares denote \textit{Scenario II}, with $P$ more than 95\%. The triangles are for a mixed regime, where the probability varies widely from 5 to 95\%. The red lines plot $2\times I_c^*$ using eq. \ref{eq:ic}. 
 (c) The sketch illustrating the relevant geometry for the bubble return.
 (d) Vertical jump velocity $\overline{u}_{I.0.5\:ms}$ averaged over the first 0.5 ms of the jump vs. $H$ for numerous bubbles. The line represents the averaged values at each $H$, completed with standard deviation.
 }
		\label{fig:PD}
\end{figure*}

To better understand under what conditions under which the return of the bubble after jump-off happens, figure \ref{fig:PD} documents the probability (\textit{P}) of return for different interelectrode distances ($H$) and as a function of $\phi$ (figure \ref{fig:PD}a) and $\overline{I}_I$ (figure \ref{fig:PD}b). 
As $H$ is increased, the transition from mode I ($P<5\%$, circles), to a mixed regime ($5\% \leq P \leq 95\%$, triangles) and finally to mode II ($P> 95\%$, squares) occurs at increasingly larger values of $|\phi|$. In fact, the dependence on $H$ is quite strong: for a fixed potential of $\phi =$ -1.3 V, $P$ changes from about 100\% at $H=117\:\mu$m to almost 0 when the distance is increased to $H=270\:\mu$m. 
The sketch in figure \ref{fig:PD}c illustrates the relevant mechanism for the bubble return. A newly formed bubble (with radius $R_0$) on one of the electrodes catches up and coalesces with the departed bubble  with radius $\hat{R}_I$. Due to momentum conservation, the resulting bubble is then located at the joint center of mass of the two coalescing bubbles, which implies a downward shift by $\Delta z$ compared to the location of the bubble with radius $\hat{R}_I$. Repeated coalescence events from both sides then bring the bubble back to the surface as seen in figure \ref{fig:sc}e. The transition between mode I and mode II is therefore governed by a competition between the departure or `jump' velocity of a bubble after coalescence and the growth rate of bubbles at the electrode. A larger magnitude of electric current, increasing approximately linearly with $\phi$ (see figure \ref{fig:dual_sum}), enables faster formation of new bubbles which then increases the likelihood of their interaction with the previously departed bubble. Upon increasing $H$ the bubble-successor needs to grow to a larger size, hence for a longer time before interacting with the already departed bubble, allowing the latter to move farther away. This will dramatically increase the current required for comeback mode.
We can capture this in a simple model based on the geometry sketched in figure \ref{fig:PD}c to predict the minimum current $I_c$ for bubble return. 
Our analysis considers the situation where the new bubble with radius $R_0$ has grown large enough to get in contact with the departing bubble. The time it takes for the bubble to grow to the radius $R_0$ is $\Delta t = kR_0^3/I_c$, where $k = \frac{8 \pi }{3} \frac{FP_g}{R_g T}$ is a prefactor containing the Faraday constant $F$, the pressure inside the bubble $P_g$, the gas constant $R_g$ and the temperature $T$ (see Supporting Information for details). During this time interval, the departing bubble travels the distance $\Delta t \cdot u_I$, with $u_I$ denoting the effective jump velocity. Based on the geometry of the triangle spanned by the centers of the two bubbles and the point A in figure \ref{fig:PD}c, the critical current for the mode transition as a function of $R_0$ is given by
\begin{equation}
    I_c(R_0;u_I,H) = \frac{u_I  R_{0}^3  k}{\left[(\hat{R}_I+R_{0})^2-\left(\frac{H}{2}\right)^2\right]^{1/2} - \hat{R}_I +R_{0}}.
    \label{eq:ic}
\end{equation}
For any $H$, a value of $R_0$ can be determined for which $I_c$ reaches a minimum value, $I_c^*$. To obtain the value of the current $I_c^*$ in this critical configuration, an estimate of the jump velocity is required. To obtain this, we tracked bubbles departing after coalescence and then averaged their vertical velocity over the first 0.5 ms to obtain $\overline{u}_{I,0.5\ ms}$. Note that $u_I$ varies widely depending on the position of both bubbles before coalescence (see Supporting Information for details).
The results for this quantity are shown in figure \ref{fig:PD}d as a function of $H$. From these data, typical values for $u_I$ are found to be in the range from 60 mm/s to 110 mm/s with a slight tendency towards higher velocities as the bubble size increases at larger electrode separations $H$. 
In figure \ref{fig:PD}b, we have included results for $2\times I_c^*$ as a function of $H$ and for different values of $u_I$. It can be seen that the model very well captures the increase of the critical current as the electrode separation increases. The best agreement between the model and the data is for $u_I = 60$ mm/s, which is close to, although slightly lower, than the measured jump velocities in figure \ref{fig:PD}d. Among potentially other factors, a reason for this slight difference is the fact that the new bubble with radius $R_0$ is also formed by coalescence and therefore also jumps off the electrode. Additionally, we do not account for shape oscillations of the larger bubbles, which become more prevalent at larger $H$.

\subsection{Performance vs. Inter-electrode distance, \textit{H}}
\begin{figure*}[h!]\centering
	\includegraphics[width=0.9\textwidth]{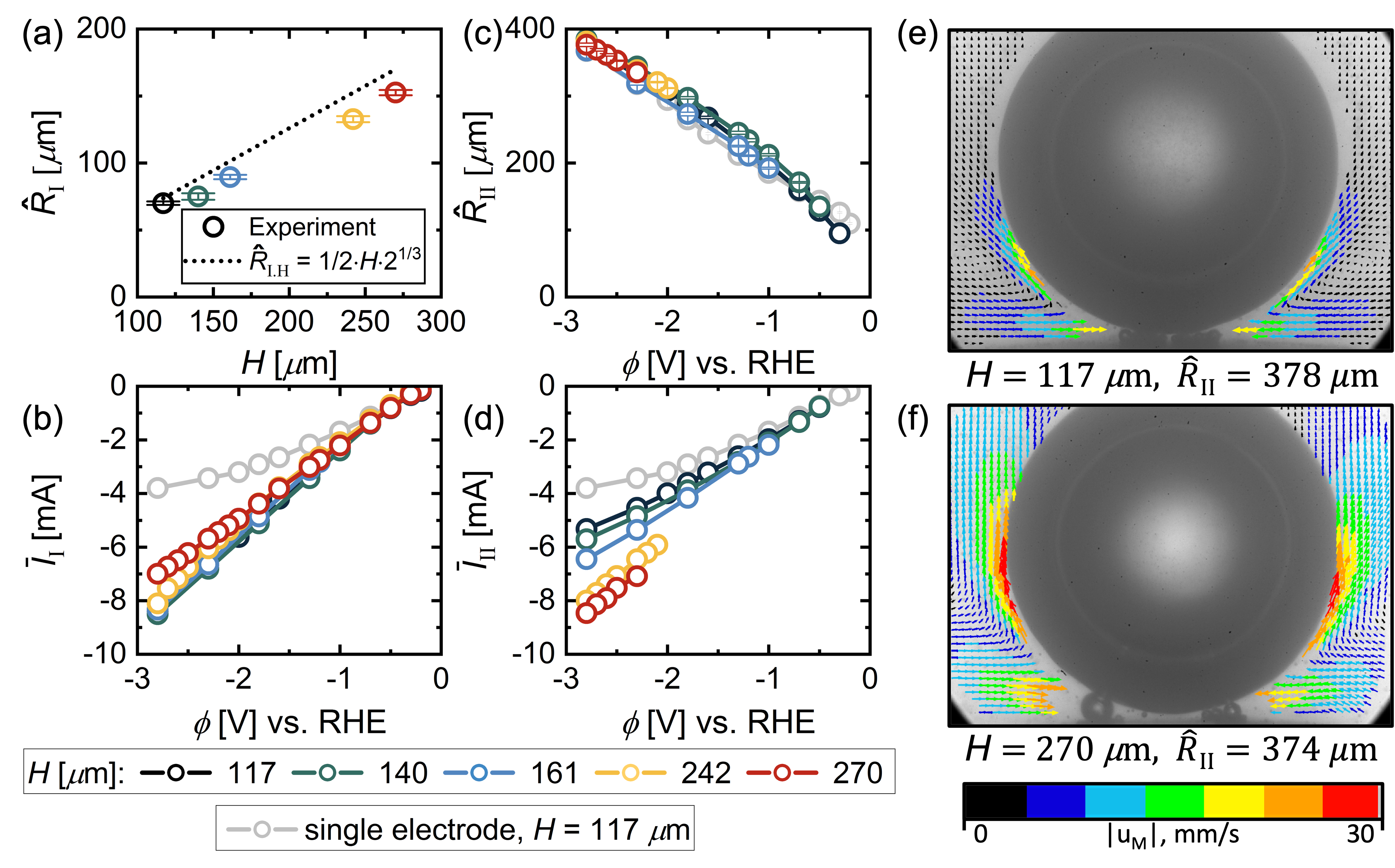}
	\caption{Departure radius (a) $\hat{R}_{I}$, (c) $\hat{R}_{II}$ and electric current (b) $\overline{I}_{I}$, (d) $\overline{I}_{II}$ for Modes \textit{I} and \textit{II}, respectively. $\hat{R}_{I}$ is given as a function of $H$. $\hat{R}_{II}$, $\overline{I}_{I}$ and $\overline{I}_{II}$ are shown as functions of potential and for various \textit{H}. Grey curves are for single electrode. (e) and (f) Velocity fields, $|u_M|$, representing Marangoni convection during mode II at -2.8 V and $H = 117$ $\mu$m and $H = 270$ $\mu$m, respectively. The velocity is measured in a period of 25 ms before the bubble departure.}
		\label{fig:dual_sum}
\end{figure*}

To understand how the current varies at different electrode separations, it is useful to first consider how the departure size of the bubbles changes for different $H$. In mode I, the departure is coalescence-driven  so that $\hat{R}_I$ is independent of $\phi$ and varies only 
 with the interelectrode distance $H$. Due to lateral oscillations of the bubble position on the electrode and possibly a slight inclination of the electrode surfaces, the results for $\hat{R}_I$ shown in figure~\ref{fig:dual_sum}a are about 10\% lower than $2^{-2/3}H$, i.e. the value for the coalescence of two bubbles each with a radius of $H/2$. This small difference was taken into account when evaluating $\hat{R}_I$ in equation~\ref{eq:ic}.

Compared to the single electrode, the current in mode I shown in figure \ref{fig:dual_sum}b is most enhanced at high overpotential and small $H$, because in this case the reduction in bubble departure size is maximal. There is only moderate decrease of $\overline{I}_I$ for larger $H$ primarily due to the relatively small range in $H$ and, consequently, in $\hat{R}_I$, which is minor compared to variations observed in $\hat{R}_s$ at different potentials. At low overpotentials, $\hat{R}_I \approx \hat{R}_s$ for the larger electrode separations studied and there is no increase of the current compared to $I_s$, just as was observed at $H = 117\:\mu$m in figure \ref{fig:sc_sum}.

In mode II, the departure radius strongly depends on the potential but at most weakly on $H$, as shown in figure \ref{fig:dual_sum}c. 
Remarkably, $\hat{R}_{II}$ is approximately the same as for the single electrode case at the same potential (see grey symbols representing $\hat{R}_s$). An investigation of the force balance \cite{thorncroft2001bubble,favre2023updated,Hossain2022} leading to these trends in $\hat{R}_{II}$ are beyond the scope of this study. Nevertheless, we present clear evidence of Marangoni convection (see figures \ref{fig:dual_sum}(e,f)), consistent with the presence of thermocapillary effects in the same potential range on single electrodes.\cite{yang2018marangoni,massing2019} Based on the flow direction, a resulting downward Marangoni force on the bubble is expected (see figure \ref{fig:scheme}a). The convective motion is much more pronounced at $H = 270\:\mu$m (figure \ref{fig:dual_sum}f) compared to the narrower spacing of $H = 117\:\mu$m in figure \ref{fig:dual_sum}e, which is in line with the difference in current between the two cases ($\overline{I}_{II} = 5.33$  mA vs. 8.46 mA, respectively). Interestingly, this does not result in a noticeable difference in $\hat{R}_{II}$ for the different interelectrode distances, which is presumably due to differences in the geometry dependent electric force.\cite{Hossain2022} 
We confirmed that the continued coalescence with small bubble does not exert a significant apparent force on the bubble (see Supporting Information for details).  

In contrast to mode I, the current in mode II shown in figure \ref{fig:dual_sum}d shows a clear dependence on the electrode separation and increases strongly for larger $H$. This is because the bubble is now centered in between the two electrodes. Therefore the electrodes become more exposed as the distance between them increases, even if the bubble size remains the same. The continuous removal of the smaller bubbles on the electrode by coalescence with the larger one proves to be very beneficial and leads to maximum currents of more than twice $\overline{I}_s$, equalling the largest currents observed in mode I.

\begin{figure*}[h!]\centering
	\includegraphics[width=1\textwidth]{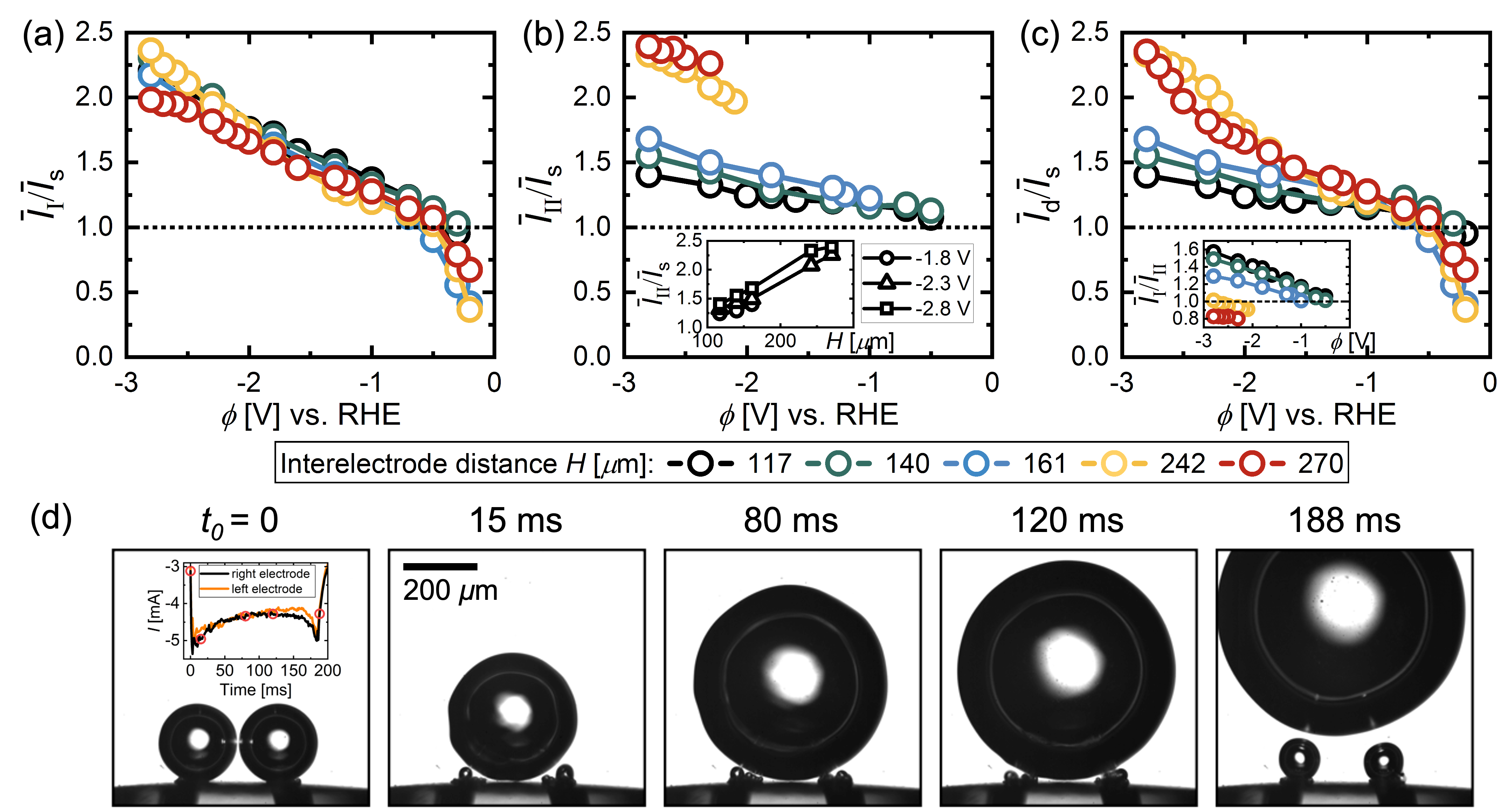}
	\caption{The electric current (a) $\overline{I}_I$, (b) $\overline{I}_{II}$ and (c) $\overline{I}_{d}$, all in dimensionless form with reference to $\overline{I}_{s}$. Data are presented as a function of potential ($\phi$) and interelectrode distance (\textit{H}). The inset in (b) shows $\overline{I}_{II}/\overline{I}_{s}$ vs. $H$ at -1.8, -2.3 and -2.8 V. The inset in (c) documents $\overline{I}_{I}/\overline{I}_{II}$ vs. $\phi$. $\overline{I}_d$ is the current averaged over both mode I ($I_I$) and mode II ($I_{II}$). (d) Snapshots throughout the bubble evolution at -2.8 V and $H=270$ $\mu$m. $t_0 = 0$ marks an instant of time one image before the coalescence of two bubbles (with radii $R_l$ and $R_r$, respectively) followed by the jump of the merged bubble off the electrode and its consecutive return. The inset shows the electric current throughout the entire evolution, with the red circles marking the corresponding snapshots.
		\label{fig:dTsum}}
\end{figure*}

To quantify the performance gain and to compensate for the $\phi$ dependence of the current, we normalise the current on the dual electrode by $\overline{I}_s$. This also accounts for small differences in $\overline{I}_s$ between the different electrodes used in this study (see Supporting Information). In figure \ref{fig:dTsum}a, the ratio $\overline{I}_I/\overline{I}_s$ is plotted for different $H$ as a function of $\phi$. As the figure shows, the interference effects at low overpotentials already discussed in the context of figure \ref{fig:sc}, cause $\overline{I}_I$ to even fall below $\overline{I}_s$ for $\phi \gtrapprox -0.5$ V. This does not improve noticeably for larger electrode spacing, presumably due to a trade-off between reduced interference effects and the increase in the bubble size with $H$. 
For larger overpotentials, the benefits of the enhanced gas removal prevail, reflected in a ratio $\overline{I}_I/\overline{I}_s>1$ which also consistently increases with increased overpotential exceeding a value of 2 at $\phi = -2.8$ V. Approximately the same values are also encountered for this potential for the ratio $\overline{I}_{II}/\overline{I}_s$ in figure \ref{fig:dTsum}b. While the performance in mode II also improves slightly for higher overpotential, it most strongly depends on $H$. As the inset in figure \ref{fig:dTsum}b shows, the ratio $\overline{I}_{II}/\overline{I}_s$ increases approximately linearly with $H$ at constant potential.

Finally, figure \ref{fig:dTsum}c shows how the resulting effective current on the dual electrode $\overline{I}_{d}$ changes relative to $\overline{I}_s$. In addition to variations in $I_I$ and $I_{II}$, this quantity is also influenced by the probability $P(H,\phi)$ of bubble return (mode II). Given the results in figure \ref{fig:PD}a, the ratio $\overline{I}_{d}/\overline{I}_s$ is therefore dominated by mode I at low and by mode II at large overpotentials. This implies that the performance gains in mode I at high $|\phi|$ are not practically realisable. However, this is only a limitation at smaller electrode separations, since the current in mode II even exceeds that of mode I for $H = 242\:\mu$m and $H = 270\:\mu$m (see inset of figure \ref{fig:dTsum}c). For these cases, the mode transition is therefore even beneficial. 

Figure \ref{fig:dTsum}d shows snapshots for the parameter combination $H = 270\:\mu$m and $\phi = -2.8$ V for which the highest ratio $\overline{I}_{d}/\overline{I}_s = 2.4$ was observed. Having the returned bubble located at the center in between the electrodes avoids the formation of larger bubbles directly on the electrodes. Notably, only a slight drop in the current is observed (see inset at $t_0= 0$) as the outline of the bubble moves beyond the electrode positions. This contradicts the common practice of considering the region under the bubble as inactive but is in line with earlier conjectures.\cite{pande2019correlating,Lake_2022}

\subsection{Conclusions}
We have explored the coalescence dynamics of electrogenerated bubbles and their influence on the electrochemical reaction rate using dual platinum micro-electrodes. We found that the coalescence of two adjacent bubbles leads to an initial jump-off of the merged bubble and premature escape from the surface. However, the continued coalescence with newly formed successors may result in a return to the electrode, hence prolonged growth. The latter mode is increasingly prevalent the higher the current and the smaller the interelectrode distance. We proposed a simple model to capture these trends and predict the critical magnitude of the current required to initiate the return process. This comeback mode negates the potential performance improvement achieved through direct departure following the coalescence event at smaller $H$ (up to a 1.7- vs. 2.3-fold increase in current at constant potential  when compared to a single electrode). However, even in cases of bubble return, the effective current at larger $H$ increased by up to 2.4 times because the bubble was then located between the electrodes, exposing a greater electrode area for the reaction. 
Therefore, this mode is promising, especially since, given the dependence on electrode separation, even greater performance gains can be expected by further increasing $H$. In practice, a similar configuration may be achieved on extended electrodes using hydrophobic islands, which should be spaced to favour coalescence-based departure and minimize the probability of bubble return, thus avoiding the blocking of the active surface area. 

\section{Conflicts of interest}
There are no conflicts to declare.

\section{Acknowledgements}
This research received funding from the Dutch Research Council (NWO) with a co-funding acquired from Nobian and Helmholtz-Zentrum Dresden-Rossendorf (HZDR) in the framework of ElectroChemical Conversion and Materials (ECCM) KICkstart DE-NL (KICH1.ED04.20.009). S.P. acknowledges the support by Basic Science Research Program through the National Research Foundation of Korea (NRF) funded by the Ministry of Education (2021R1A6A3A14039678). D.K., D.L. and M.T.M.K. received funding from the European Research Council (ERC) (BU-PACT grant agreement number 950111, ERC Advanced grant number 740479-DDD and ERC Advanced Grant ‘FRUMKIN’ number 101019998, respectively). We thank V. Sanjay for insightful discussions on the subject.

\bibliography{1_references}
\end{document}


\newpage
\section{Single electrode: characterization}

\begin{figure*}[h!]\centering
	\includegraphics[width=0.9\textwidth]{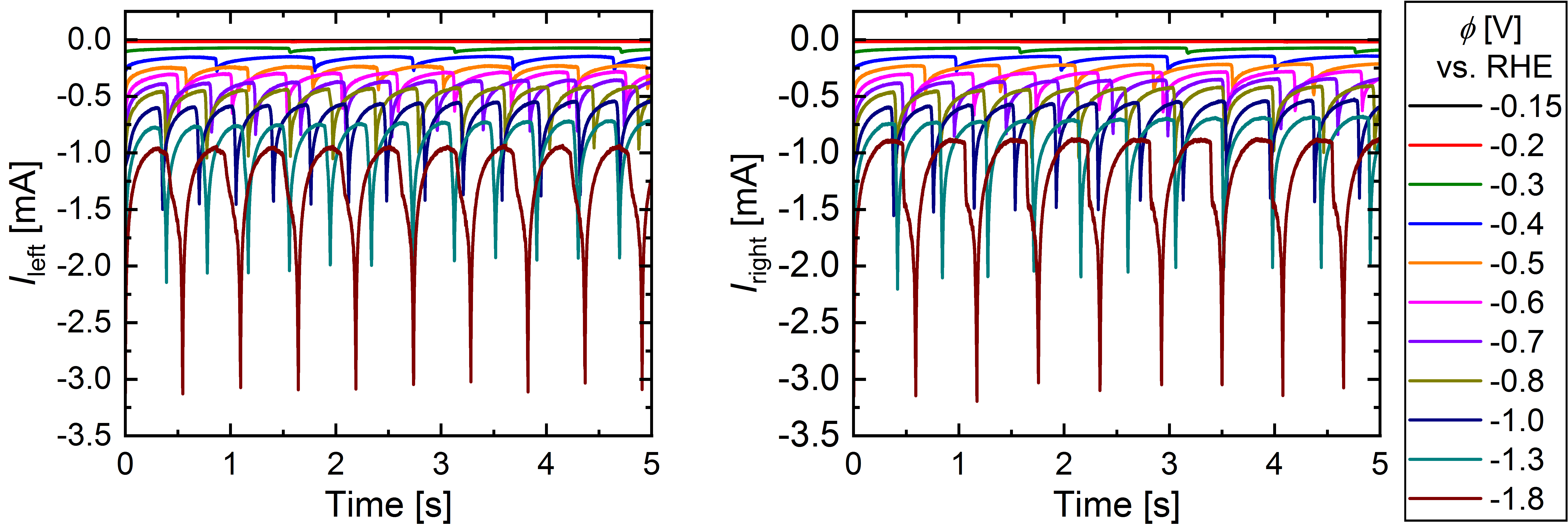}
	\caption{The electric current over time for left and right electrodes.
		\label{ESI:single}}
\end{figure*}

Figure \ref{ESI:single} documents the electric current over 5 seconds (out of 30 seconds) of the experimental run at various potentials ($\phi$). The currents are shown for $H=117\:\mu$m for left and right electrodes (see figure 1a in the manuscript), run separately. 

\begin{figure*}[h!]\centering
	\includegraphics[width=0.9\textwidth]{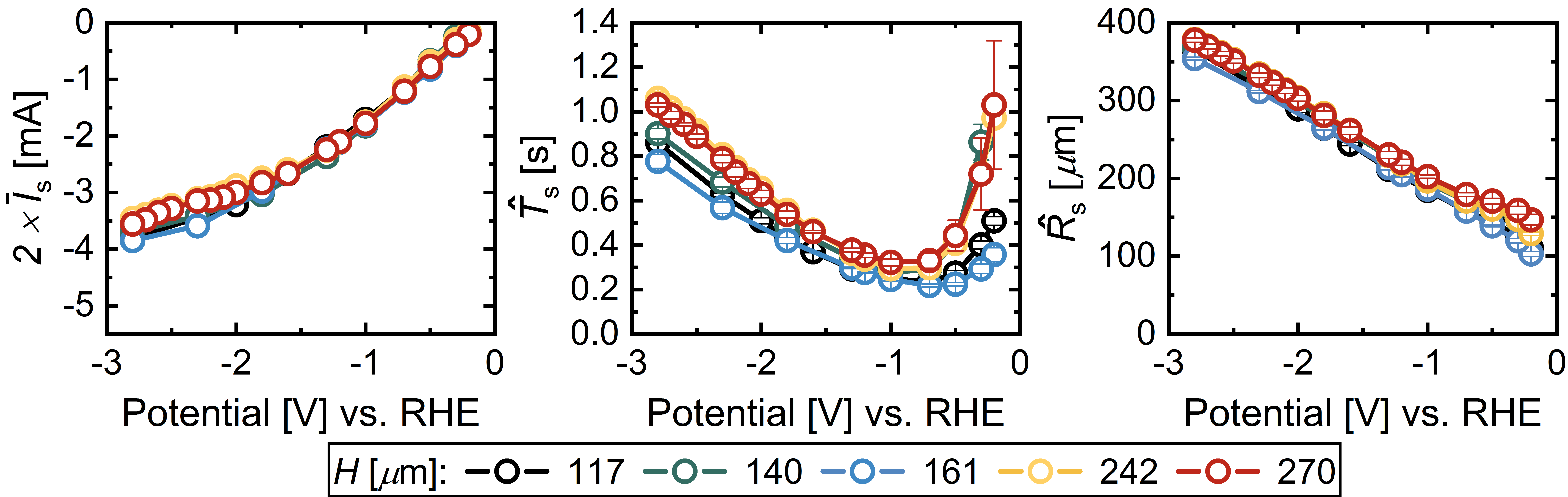}
	\caption{The electric current, lifetime and departure radius for H$_2$ bubbles produced at single electrode vs. $\phi$ for different electrodes ($H$). The error bars represent standard deviation.
		\label{ESI:single2}}
\end{figure*}

Figure \ref{ESI:single2} characterizes five electrodes in terms of the electric current ($2 \times \overline{I}_s$), lifetime ($\hat{T}_s$), and radius at the departure ($\hat{R}_s$) vs. $\phi$. $\overline{I}_s$ is averaged over 30 seconds and the left and right electrodes. $\hat{T}_s$ and $\hat{R}_s$ are averaged for multiple bubbles and accompanied by the standard deviations. 
The low standard deviation for $\hat{R}_{s}$ and $\hat{T}_{s}$ demonstrates the highly periodical evolution of bubbles. The quite similar current between various $H$ suggests that the surfaces of these electrodes are alike. However, the small differences are enough to affect the dynamics of H$_2$ bubbles and significantly alter the lifetime and size at the departure.

\section{Dual electrode: characterization}
\begin{figure*}[h!]\centering
	\includegraphics[width=1\textwidth]{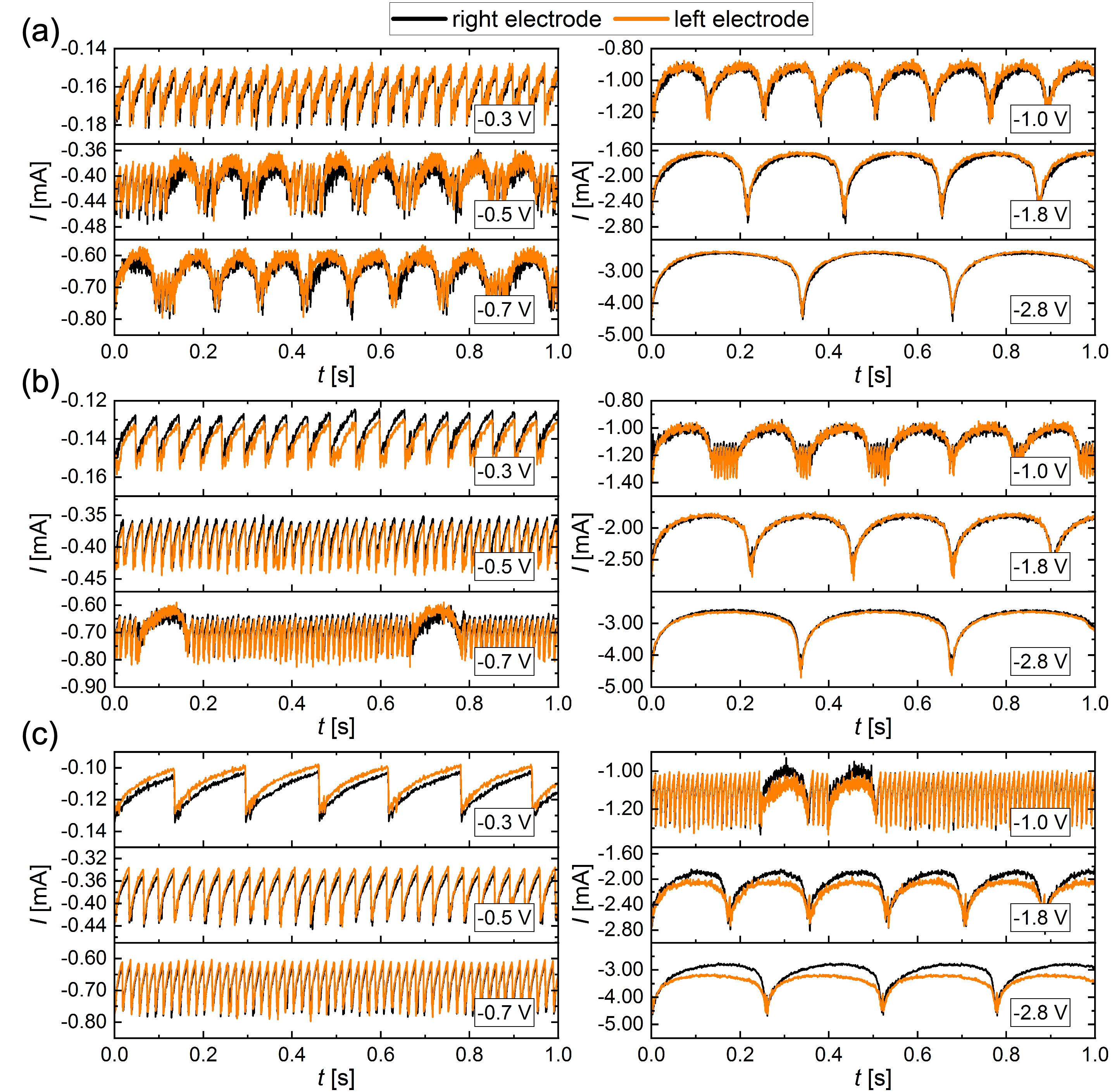}
	\caption{The electric current vs. time plotted for 1 second out of 30 seconds of the experimental run for various potentials $\phi$ and interelectrode distance: (a) $H=117\:\mu$m, (b) $H=140\:\mu$m, (c) $H=161\:\mu$m.
		\label{ESI:curr1}}
\end{figure*}

\begin{figure*}[h!]\centering
	\includegraphics[width=1\textwidth]{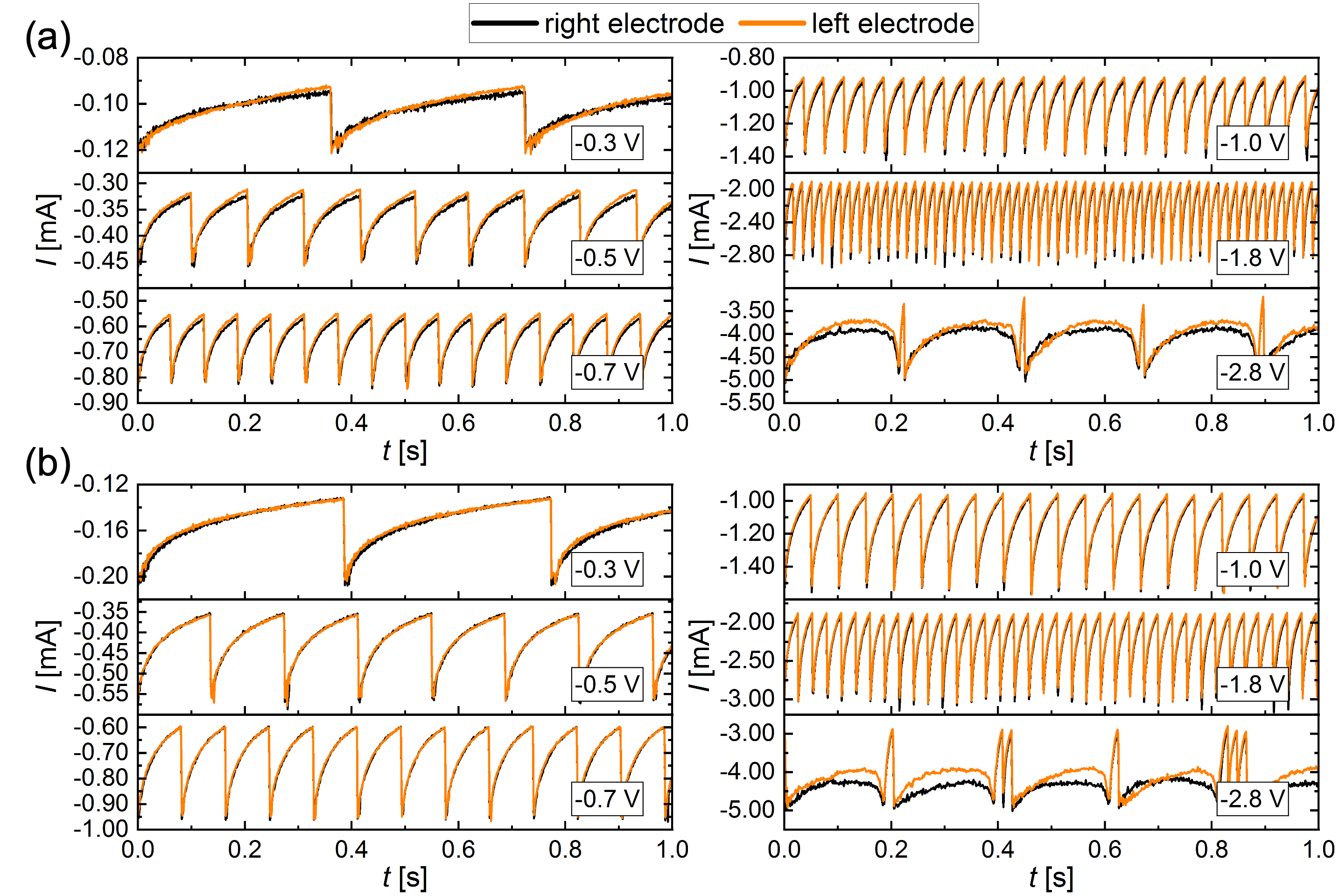}
	\caption{The electric current vs. time plotted for 1 second out of 30 seconds of the experimental run for various potentials $\phi$ and interelectrode distance: (a) $H=242\:\mu$m, (b) $H=270\:\mu$m.
		\label{ESI:curr2}}
\end{figure*}

For completeness of the results presented in the manuscript figures \ref{ESI:curr1} and \ref{ESI:curr2} document the electric current ($I$) vs. time plotted for 1 second our of 30 seconds of the experiment run. $I$ is plotted for various potentials ($\phi$) and interelectrode distance ($H$).
Figs. \ref{ESI:curr1}a, b and c are for $H=117\:\mu$m, $H=140\:\mu$m, $H=161\:\mu$m, respectively. Figs. \ref{ESI:curr2}a and b are for $H=242\:\mu$m and (b) $H=270\:\mu$m.

\begin{figure*}[h!]\centering
	\includegraphics[width=0.6\textwidth]{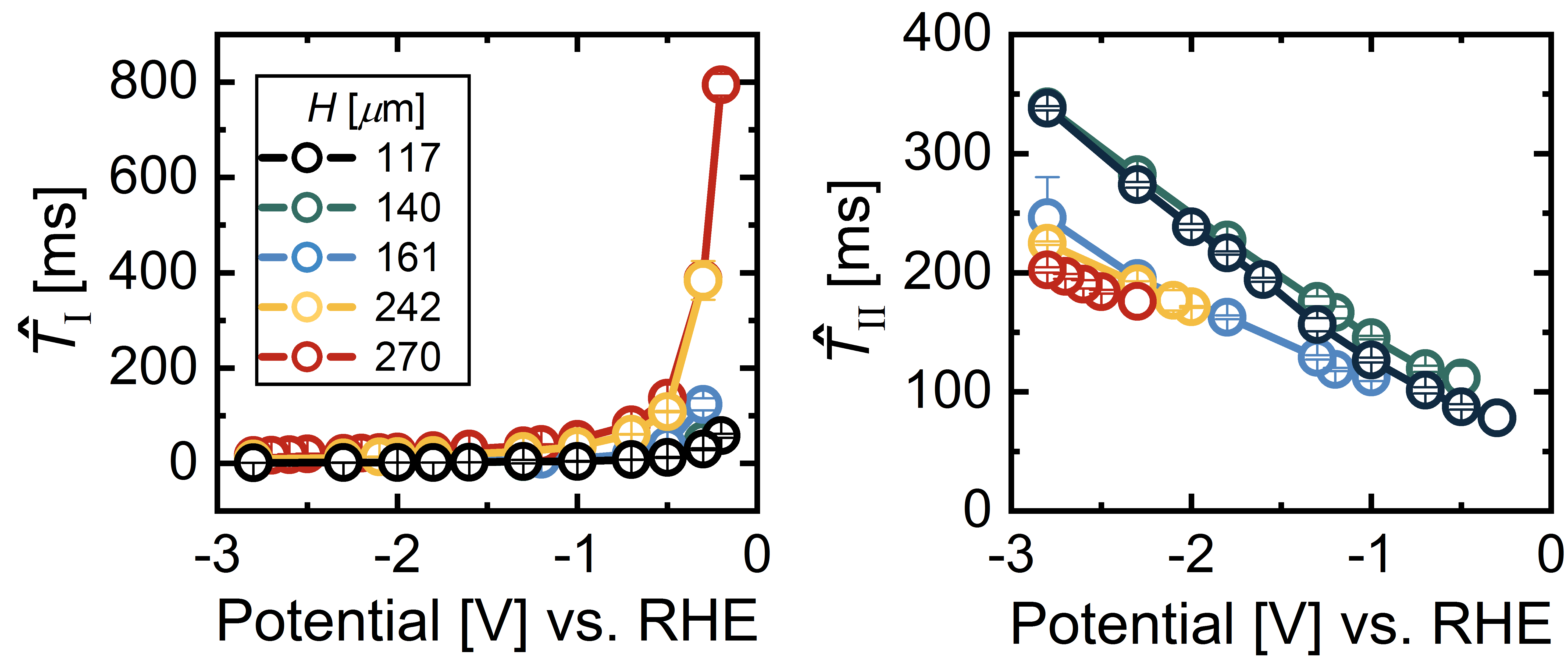}
	\caption{The lifetime at dual electrode as a function of potential ($\phi$) and interelectrode distance ($H$). $\hat{T}_I$ and $\hat{T}_{II}$ are for mode I and mode II, respectively. The error bars represent standard deviation.
		\label{ESI:dual1}}
\end{figure*}

Figure \ref{ESI:dual1} documents the lifetime of the bubbles produced at dual electrode vs. potential ($\phi$) for Modes I (left) and II (right) and for different electrodes. $\hat{T}$ is averaged for multiple bubbles and accompanied by the standard deviations. Two main trends can be observed: (i) Since the departure radius in mode I is independent of the $\phi$, $\hat{T}_I$ reduces at larger overpotentials, owing to higher electric current. It also increases together with $H$, especially at larger $\phi$. --- This is because the pair of bubbles need to grow to a bigger size before coalescence at larger $H$; (ii) On the other hand, $\hat{T}_{II}$ increases at larger overpotentials and reduces at larger $H$. As already mentioned in the manuscript, larger overpotentials imply larger downward-acting forces increasing the departure size of the bubble. Therefore the bubble would grow for a longer time. However, the separation of two electrodes away from each other for larger distances ($H$) enables higher currents, hence faster H$_2$ production rate.

\begin{figure*}[h!]\centering
	\includegraphics[width=0.5\textwidth]{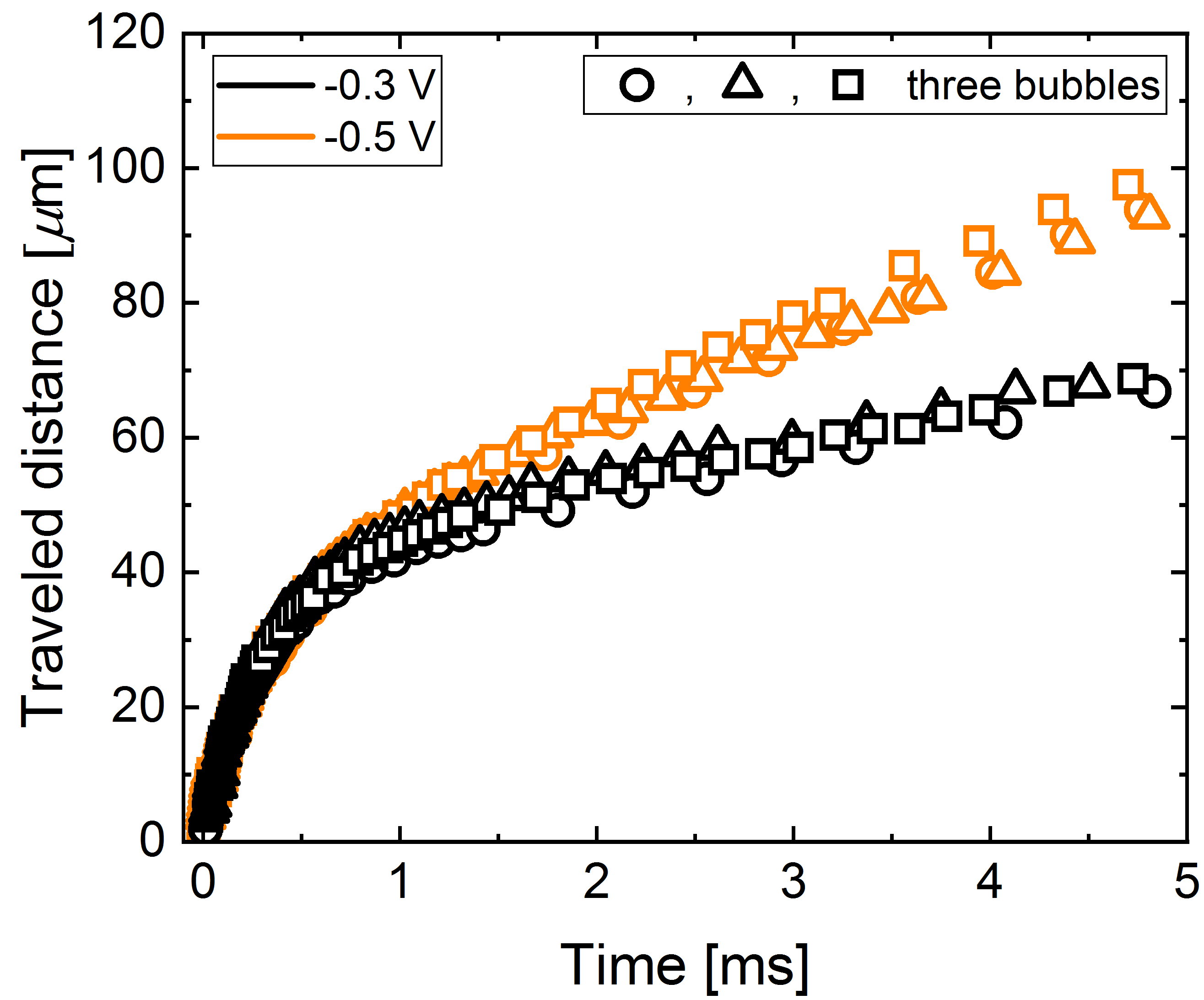}
	\caption{The trajectory of bubble at -0.3 V and -0.5 V over the first 5 milliseconds after coalescence driven jump-off. $t_0=0$ is one frame before coalescence. Each potential is presented by three bubbles. The interelectrode distance $H = 117$ $\mu$m.
		\label{ESI:dual2}}
\end{figure*}

Figure \ref{ESI:dual2} demonstrates the traveling distance in vertical direction of the merged bubble in the first 5 milliseconds after the jump-off of the electrode driven by the coalescence event at $H = 117\:\mu$m. The results are shown for $\phi = -0.3$ and $-0.5$ V and three bubbles in each case. The curves document that the jumping velocity notably decays over time ($\Delta t$) due to the effects of drag force, particularly in the moments following the jump. Interestingly, the "terminal" velocity of the bubble, which can be read from the slope of the curves and i.e. ca. after 2 ms, increased at a larger overpotential (-0.5 V). This might be explained by the wake behind the rising bubble-predecessor. This flow drags, hence accelerates the merged bubble in the moment of departure. The wake enhances with the faster departure of bubbles-predecessors (smaller $T$) which is the case at a more negative potential.

\begin{figure*}[h!]\centering
	\includegraphics[width=1\textwidth]{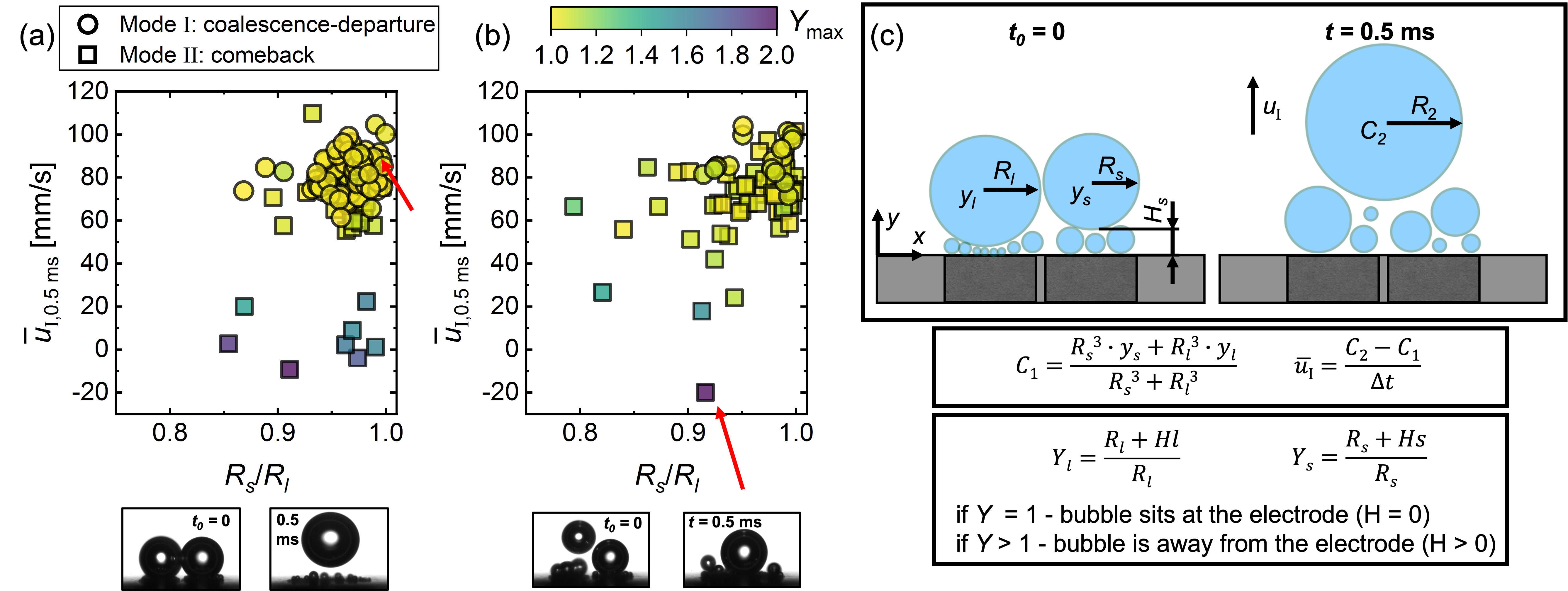}
	\caption{(a),(b) The vertical jumping velocity ($\overline{u}_{I,0.5ms}$) 
 vs.. $R_s/R_l$ at $\phi = -0.5$ V and $\phi = -1.0$ V. The color bar scales the relative position of the bubble shown in (c), i.e. the maximum of either smaller or bigger bubbles, prior to coalescence at $H = 117$. Image recording performed at 10 kHz.
 \label{ESI:dual3}}
\end{figure*}

Figure \ref{ESI:dual3} represents the vertical jumping velocity $\overline{u}_{I.0.5\:ms}$ for numerous bubbles, averaged over the first 0.5 ms of the jump, vs. parent size ratio $R_s/R_l$. The experiments performed at $H = 117$ and (a) $\phi = -0.5$ V, (b) $\phi = -1.0$ V. $R_s$ and $R_l$ are radii for smaller and larger bubbles, respectively. The geometric parameters are shown in figure \ref{ESI:dual3}c. The color bar scales another geometric parameter $Y_{max}$ given in dimensionless form. It represents the relative position of the bubble i.e. the distance from the bubble bottom to the electrode, chosen as the maximum value between the smaller and bigger bubbles. The bubble sits at the electrode if $Y = 1$ and is away from the electrode if $Y>1$. The circles represent mode I, i.e. when the bubble departs into the bulk following the coalescence event, and squares denote mode II, i.e. when the once departed bubble following the coalescence event comes back to the surface, continues to grow, and departs at a later stage due to buoyancy. Note that $u_I$ varies widely depending on the position of both bubbles before coalescence and their size ratio. 

Figures \ref{ESI:dual3}a and b demonstrate two general trends: (i) $\overline{u}_{I,0.5ms}$ reduces as per reduction in $R_s/R_l$ and as per increase in $Y_{max}$; (ii) at higher potential, hence current, the bubble would come back to the electrode more often moving away with even higher jumping velocity, as already emphasized in the manuscript (see figure 5).

Upon coalescence of two bubbles, there is a release of surface energy ($\Delta G_s$) given as
\begin{equation}
    \Delta G_s = 4\pi\gamma(R_l^2+R_r^2-R_I^2),
     \label{ESI:eq1}
\end{equation}
where $R_l$, $R_r$ and $R_I$ are the left, right and merged bubbles, respectively. $\gamma \approx 0.072$ N/m is the surface tension of the electrolyte. The released energy partly dissipates by the bubble oscillations, working against viscous drag. When in the proximity to the surface, the remaining energy is converted to the kinetic energy ($E_k$) driving the resultant (merged) bubble to jump off the electrode.\cite{lv2021self} The kinetic energy is

\begin{equation}
    E_k = \frac{1}{2}C_M\rho_l\frac{4}{3}\pi R_I^3 u_I^2,
    \label{ESI:eq2}
\end{equation}
where $C_M$ is added mass coefficient, $\rho_l$ electrolyte density and $u_I$ is the initial jumping velocity. For a spherical bubble $C_M = 0.5$, however, when the bubble is in proximity to the wall the coefficient is larger.\cite{raza2023coalescence}
In detail, when two bubbles approach each other, the thin film of electrolyte separating them gradually drains $\BigOSI[\big]{}{\mu s}$ and eventually ruptures. This leads to the formation of a neck, i.e. an open cavity, and a series of capillary waves of varying strengths that propagate along the electrolyte-gas interface. These waves move away from the neck region until they meet at the opposite apex of the coalescence point (see manuscript, fig. 3c).
The strength of these waves decreases as they travel along the interface due to continuous viscous dissipation\cite{sanjay2021bursting}. 
Meanwhile, the surface tension $\gamma$ drives the retraction of the remaining capillary waves towards a spherical shape, deforming the bubble shape. 
Once the excess surface energy overcomes the work done by the bubble against viscous drag ($W_\mu$) during the expansion and retraction processes, the resultant net component of momentum perpendicular to the surface causes the bubble to jump off the electrode. As neither of the bubbles is attached to the electrode, the adhesion energy $W_a$ is neglected. The process is controlled by surface tension and viscosity and is often characterized in terms of the Ohnesorge number ($Oh=\frac{\mu}{\sqrt{\rho \gamma R_I}}$)\cite{sanjay2021bursting}. $\mu$ is the dynamic viscosity of the electrolyte. The influence of gravity during the coalescence, before lift-off, is negligible\cite{raza2023coalescence}. While the process is considered highly inefficient, with only a small portion of surface energy translating into kinetic energy \cite{raza2023coalescence}, it is sufficient for a resultant bubble ($R_I$) to jump off the electrode acquiring an initial velocity $u_I$. 

\begin{figure*}[h!]\centering
	\includegraphics[width=1\textwidth]{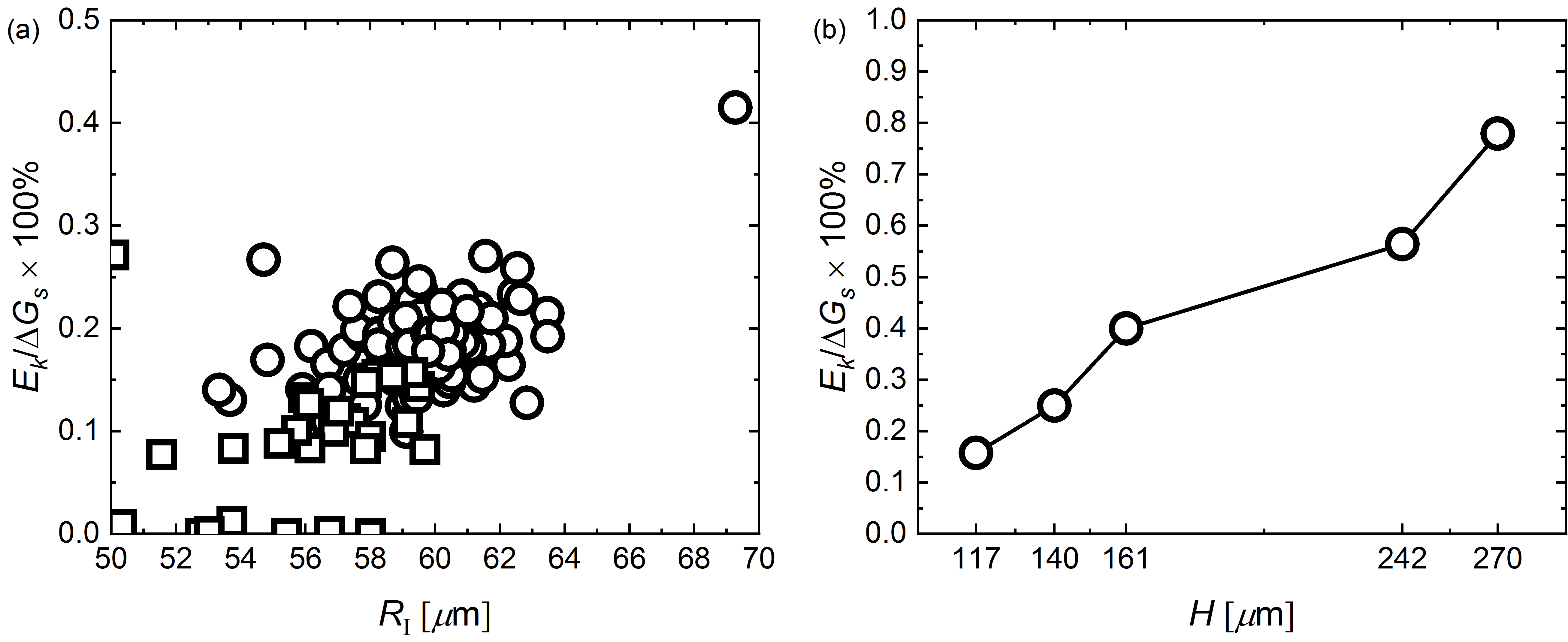}
	\caption{The ratio between the translated kinetic energy ($E_k$) and the released surface energy ($\Delta G_s$) as a function of (a) $R_I$ and (b) $H$.
		\label{ESI:dual4}}
\end{figure*}

Figure \ref{ESI:dual4} documents an estimation of the ratio between the translated kinetic energy ($E_k$) and the released surface energy ($\Delta G_s$) as a function of (a) $R_I$ and (b) $H$. The data points in fig. \ref{ESI:dual4}a is calculated using eq. \ref{ESI:eq1} and eq. \ref{ESI:eq2} by taking corresponding $R_l$, $R_r$, $R_I$ and $\overline{u}_{I,0.5 ms}$ for the bubbles presented in fig. \ref{ESI:dual3}a, i.e. at $H=117\:\mu$m and $\phi = -0.5$ V. The circles represent the bubbles following mode I and squares mode II. The data points in fig. \ref{ESI:dual4}b is based on the $\hat{R}_I$ from fig. 6a and $\overline{u}_{I,0.5ms}$ from fig. 5d in the manuscript, assuming that $R_l = R_r$, hence $R_l = \frac{\hat{R}_I}{2^{1/3}}$. 

Although the taken velocity $\overline{u}_{I,0.5ms}$ is smaller than the initial jumping velocity and is further affected by the wake behind the departed bubble-predecessor, fig. \ref{ESI:dual4} confirms the inefficiency of the coalescence process, as the translated kinetic energy is below 1\% of the released surface energy.

\section{Phase diagram: model}
\label{ESI:model}
To provide a general understanding of the relationship between probability ($P$), interelectrode distance ($H$), and electric current ($I$) presented in fig.~5 (see manuscript), we propose a simple model based on the velocity of the departing bubble and the growth rate of the bubble-successor. The model predicts the critical magnitude of the current $\overline{I}_{c}$ required to produce hydrogen quickly enough to guarantee the coalescence between the departing bubble ($\hat{R}_I$) and its successor ($R_0$). The time it takes for the new bubble to grow to the radius $R_0$ is $\Delta t$ which can be found from Faraday's and ideal gas laws.

The ideal gas law given as

\begin{equation}
\label{ideal1}
    P_gV = \frac{m}{M}R_g T,
\end{equation}
where $P_g$ is the pressure inside the bubble, $V$ is the volume of the gas, $m$ is the total mass of the gas, $M$ is the molar mass, $R_g$ is the ideal gas constant and $T$ is temperature of the gas.
The produced charge is

\begin{equation}
\label{charge1}
    Q = \overline{I}_c \cdot \Delta t.
\end{equation}
The Faraday's law reads

\begin{equation}
\label{Faraday1}
    \frac{m}{Q} = \frac{1}{F} \frac{M}{\nu},
\end{equation}
where $F$ is the Faraday's constant and $\nu$ is the valency of the ions (+1) multiplied by number of protons (2).

By substituting Eqs. \ref{charge1} and Eq. \ref{Faraday1} in Eq. \ref{ideal1}

\begin{equation}
\label{3}
    \Delta t = \frac{R_0^3}{\overline{I}_c} \cdot \frac{8 \pi}{3} \frac{F P}{R_g T} = \frac{R_{0}^3}{I_c}\cdot k,
\end{equation}
with $k = \frac{8 \pi}{3} \frac{F P}{R_g T}$. During this time interval $\Delta t$, the departing bubble travels the distance $\Delta t \cdot u_I$, with $u_I$ denoting the effective jump velocity. Based on the geometry of the triangle spanned by the centers of the two bubbles and the point A in figure 5c (see manuscript), the following relation can be expressed:


\begin{equation}
\label{4}
    \left[(\hat{R}_I+u_I \cdot \Delta t)-R_0\right]^2+\left(\frac{H}{2}\right)^2 = (\hat{R}_I+R_0)^2
\end{equation}
where $\hat{R}_I = 2^{1/3}\cdot \frac{H}{2}$ or is taken from experiment (see figure 6a).

By substituting $\Delta t$ into eq. \ref{4}, the critical current for the mode transition as a function of $R_0$ is given by:

\begin{equation}
    I_c(R_0;u_I,H) = \frac{u_I  R_{0}^3  k}{\left[(\hat{R}_I+R_{0})^2-\left(\frac{H}{2}\right)^2\right]^{1/2} - \hat{R}_I +R_{0}}.
    \label{eq:ic_ESI}
\end{equation}

\section{Apparent force: bubble-carpet interaction}
It was shown that the H$_2$ bubble grows while seated on a carpet of microbubbles, undergoing a continuous intensive coalescence with it.
Throughout numerous coalescence events, the mother bubble experiences a shift in the mass center, resulting in acceleration towards the electrode surface. This acceleration can be considered as an apparent force $F_{g-g}$.
An estimate for  this apparent force is derived from the principle of momentum conservation, given as

\begin{equation}
\label{N1}
    m c_M = m_0c_0 + \dot{m} t c, 
\end{equation}
where $m$ and $c_m$ represent the total mass and mass center of the bubble at time $t$, respectively; $m_0$ and $c_0$ denote the initial mass and mass center of the bubble at an arbitrary time $t_0 = 0$; and $\dot{m}$ stands for the mass injection rate due to continuous coalescence with the carpet of microbubbles with the center of mass at $c$.  $m_0$ and $c_0$ are constant. All mass centers are with reference to the electrode surface. The mass of the bubble $m$ at any given moment is

\begin{equation}
\label{N2}
    m = m_0 + \Dot{m}t.
\end{equation}

By substituting $\Dot{m}t = m-m_0$ from eq. \ref{N1} into eq. \ref{N2} 

\begin{equation}
\label{N3}
   m c_M = m_0c_0 + m c - m_0c.
\end{equation}

Given that $c$ at a given current $I$ is a constant (as well as $m_0$ and $c_0$)

\begin{equation}
\label{N4}
    m \dot{c_M} = \dot{m}(c-c_M).
\end{equation}

We further assume that $\dot{m}$ is also constant at given $I$. By performing differentiation one more time using eq. \ref{N4}

\begin{equation}
\label{N5}
    m \ddot{c_M}= -2 \dot{m}\dot{c_M}.
\end{equation}

By substituting eq. \ref{N4} in eq. \ref{N5}

\begin{equation}
\label{N6}
    F_{g-g} = m \ddot{c_M}= -2 \frac{\dot{m}}{m}\dot{m} (c-c_M).
\end{equation}

When considering only a short interval $t$,  we can approximate $c_M \approx R$ and eq. \ref{N6} reads

\begin{equation}
\label{N7}
    F_{g-g} = 2 \frac{1}{m}(\frac{m}{t})^2 R.
\end{equation}
$c$, which is the thickness of the carpet, is also neglected. Using the Faraday's and ideal gas laws

\begin{equation}
\label{N8}
    \dot{m} = \frac{m}{t} = \frac{\rho_g V}{V} = \frac{\rho_gItk}{t} = \rho_g I k,
\end{equation}
where $k = \frac{R_g T}{2P_g F}$.

Finally, by substituting eq. \ref{N8} into eq. \ref{N7} the apparent force reads

\begin{equation}
\label{N9}
    F_{g-g} = \frac{3}{2} \frac{\rho_g I^2 k^2}{\pi R^2}.
\end{equation}

\begin{figure*}[h!]\centering
	\includegraphics[width=0.5\textwidth]{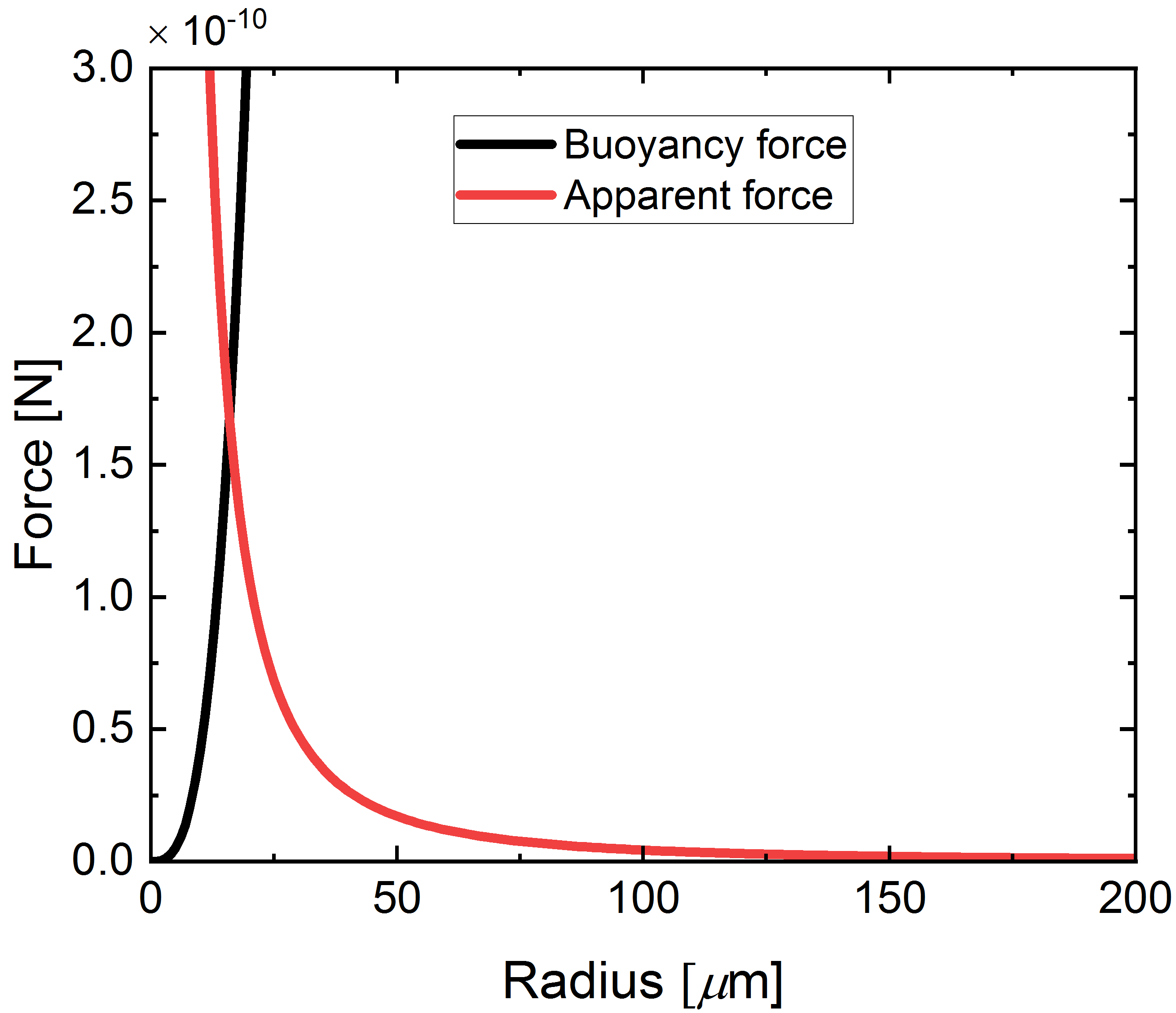}
	\caption{Buoyancy and apparent forces as a function of bubble radius at $I = - 8$ mA.
		\label{ESI:app}}
\end{figure*}

Figure \ref{ESI:app} documents a comparison between the buoyancy force and apparent force, calculated using eq. \ref{N9} at $I = -8$ mA (which corresponds to approximately the maximum current observed in this study), against the bubble radius $R$. The buoyancy force defined as $F_b = (\rho_g - \rho_{H2}) g V$, where $g$ represents gravitational acceleration. Fig. \ref{ESI:app} demonstrates that the continued coalescence with the carpet of microbubbles does not impose a significant apparent force on the bubble. --- This force decays rapidly with increasing $R$ and becomes smaller than the buoyancy force at at approximately $R = 16\:\mu$m.

\bibliography{2_supp}